# Scale-free flocking and giant fluctuations in epithelial active solids

Yuan Shen[1]*, Jérémy O'Byrne[2], Andreas Schoenit[1], Ananyo Maitra[2,3], Rene-Marc Mege[1], Raphael Voituriez[2]*, and Benoit Ladoux[1,4,5]*

[1] Universite Paris Cite, CNRS, Institut Jacques Monod, F-75013 Paris, France.
[2] Laboratoire Jean Perrin, CNRS, Sorbonne Université, Paris, France
[3] LPTM, CNRS/CY Cergy Paris Université, F-95032 Cergy-Pontoise cedex, France
[4] Department of Physics, Friedrich-Alexander Universität Erlangen-Nürnberg, Erlangen, Germany
[5] Max-Planck-Zentrum für Physik und Medizin, Erlangen, Germany

* Correspondence: yuan.shen@ijm.fr, raphael.voituriez@sorbonne-universite.fr and benoit.ladoux@fau.de

Classification: Physical Science, Biophysics and Computational Biology

**Abstract**

The collective motion of epithelial cells is a fundamental biological process which plays a significant role in embryogenesis, wound healing and tumor metastasis. While it has been broadly investigated for over a decade both *in vivo* and *in vitro*, large scale coherent flocking phases remain underexplored and have so far been mostly described as fluid. In this work, we report a mode of large-scale collective motion for different epithelial cell types in vitro with distinctive new features. By tracking individual cells, we show that cells move over long time scales coherently not as a fluid, but as a polar elastic solid with negligible cell rearrangements. Our analysis reveals that this solid flocking phase exhibits signatures of long-range polar order, unprecedented in cellular systems, with scale-free correlations, anomalously large density fluctuations, and shear waves. Based on a general theory of active polar solids, we argue that these features result from massless Goldstone modes, which, in contrast to polar fluids where they are generic, require the decoupling of global rotations of the polarity and in-plane elastic deformations in polar solids. We theoretically show and consistently observe in experiments that the fluctuations of elastic deformations diverge for large system size in such polar active solid phases, leading eventually to rupture and thus potentially loss of tissue integrity at large scales.

**Keywords:** active solid, epithelial cell, collective motion

**Significance statement**

During embryonic development and wound healing, epithelial cells usually display in-plane polarity over large spatial scales and move coherently. However, over years, most *in vitro* studies have examined the fluid-like chaotic dynamics of epithelial cells, in which collective cellular flows self-organize into recurring transient vortices and jets similar to those observed in classical fluid turbulence. Little is known about the large-scale coherent dynamics of epithelial cells. We demonstrate that such coherent motions are not simply turbulent-like flows with larger correlation lengths, but a new mode of collective motion with a solid-like behavior, accompanied by an emergent global order, scale-free correlations, anomalous density fluctuations and propagating Goldstone modes. Our work suggests that such a collective motion of epithelial cells falls outside the scope of traditional active fluids, which may shed new light on the current studies of collective cell migration as well as active matter physics.

**Introduction**

Epithelia are confluent, protective layers of cells that cover organs throughout animal bodies. The collective motion of epithelial cells is of great importance since it contributes to numerous fundamental biological processes, including embryogenesis, wound healing, and tumor metastasis (1). Previous *in vitro* studies have shown that 2-dimensional epithelial cell monolayers can exhibit different modes of collective motion qualitatively comparable to fluid turbulence, with velocity correlations that vanish at large scales (2-10) and thus, with no global coherent motion for large system sizes. The onset of such flowing phases was found to depend on different cellular parameters, such as cell density (11, 12), cell/cell adhesion (13, 14), or single cell motility (15, 16) as well as external conditions, including substrate stiffness (6, 17), boundary effects (18), and surface curvatures (19). At the theoretical level, these observations motivated the development of different classes of physical models (20-26) that consistently put forward an analogy with the unjamming transition in colloidal glasses, which separates a static amorphous solid phase and a flowing liquid phase; following this analogy, cellular flows have, till now, been usually associated to a fluid-like rheological description of epithelia.

More recently, a growing body of evidence suggests that epithelial cells can also move coherently over large scales (24, 27-30), while remaining in a fluid phase, in a migration mode qualitatively comparable to flocking behaviors that have been reported in various living or artificial systems consisting of an assembly of interacting self-propelled agents, such as bird flocks or self-propelled colloids (31, 32). On the theoretical side, such 2-dimensional flocking phases with long-range order – i.e. non-vanishing velocity correlations for arbitrary large systems – are a hallmark of active systems because they have no equivalent in equilibrium materials (33, 34). Such flocking phases were predicted theoretically and shown to arise in minimal models of active systems, where agents are free to move and are neither repulsive nor cohesive, from the conspiracy of short-ranged alignment interaction and self-propulsion (34, 35).

However, in many experimental systems, and in particular epithelial systems, as the density of active agents or attractive interactions keep increasing, the description of free-to-move agents, thus remaining in a liquid phase, does not hold anymore and solid phases with conserved local structures are expected. This naturally called for the study of active solids (36-38). Despite the extensive research on active fluids, the study of active solids remains largely unexplored, especially in biological systems. In particular, in the context of cellular systems so far most of observations of solid phases were interpreted as jammed non-motile states where cells were considered to be caged by their neighbors and behave like glass (12, 21, 39), with the exception of a recent work by Lang et al. (40). On the other hand, theoretical modeling showed that cell tissues could in principle flock as solids (24, 41, 42) which, however, has not been experimentally demonstrated. Some recent observations of seemingly unjamming transitions of epithelial tissues showed that cells could move coherently on a relatively large scale at high densities (7, 27, 43), but the physical properties of such flocking dynamics, and in particular their fluid or solid rheological properties remain to be elucidated.

In this work, we provide a systematical experimental and theoretical investigation of the long-range coherent motion of skin epithelial cells. Using these epithelial monolayers, we uncover the emergence of large-scale collective migration modes, featuring new physical characteristics. By tracking individual cells, we demonstrate that cells move coherently over long timescales as a polar elastic solid with minimal cell rearrangements, in contrast to the earlier descriptions of fluid flocking phases. Our analysis uncovers that this solid flocking phase exhibits striking signs of long-range polar order, with scale-free correlations, anomalously large density fluctuations, and shear waves. On the basis of a general theory of active polar solids, we propose that these features arise from massless Goldstone modes associated with broken rotation symmetry. Unlike in polar fluids where they are expected to be generic, these modes in solids require the decoupling of global polarity rotations from in-plane elastic deformations. Both theoretical predictions and experimental observations consistently show that in such polar active solid phases, elastic deformation fluctuations increase with system size, eventually leading to rupture, which could imply in biological systems loss of tissue integrity at larger scales.

**Results**

**Synchronization and solid-like dynamic behavior.** To induce the solid-like long-ranged collective motion, skin epithelial cells (HaCaT) are initially cultured in a serum deprived state for 2 days and then stimulated by adding fetal bovine serum (the measurements start around 6 hours after stimulation, see *Methods* for details) using a previously established method (28). After around 6 hours, cells are awakened from the static quiescent state and start moving (28). Initially, cells move randomly, but they soon synchronize their motions (Movie S1). The average speed of the cells, $<|v|>$, increases gradually with time from ~0.05 to ~0.3 μm min$^{-1}$ within 1500 min (Fig. 1A). It then reaches the maximum of ~0.45 μm min$^{-1}$ at $t$ = ~1900 min and then continuously decreases to ~ 0.05 μm min$^{-1}$ after ~700 min. Simultaneously, the velocity order parameter, $S_v = \left|\sum_i^N \mathbf{v}_i\right|/\sum_i^N |v_i|$ (where $N$ is the total number of cells in the field of view, $\mathbf{v}_i$ and $v_i$ are the velocity and speed of individual cells, respectively) gradually increases from ~0 to ~0.8 within 1500 min and then saturates at ~0.95 for ~500 min, clearly indicative of coherent collective motion, or flocking phase. At later times, $S_v$ suddenly decreases, and the sample eventually gets into a jammed state where the cells are caged by their neighbors and jiggle locally. It is noted that $S_v$ forms a valley at $t$ = ~2250 min. This is because cells change their moving direction due to the finite size of the sample dish. To determine the rheological properties of the epithelium in the flocking phase, we record the motion of cells for 3 hours at a frame rate of 10 min per frame and then define the cells which change their neighbors more than 5 times during this process as "floppy cells". We compare the dynamics of HaCaT cells with MDCK cells, originating from kidney, where a turbulent fluid-like collective dynamic behavior is well established (6, 12). We find the floppy cells form large clusters and percolate throughout the whole MDCK monolayer (Fig. 1D) but only form small clusters in HaCaT cells (Fig. 1C). In addition, Fig. 1B characterizes how the average distance between cells and their initial neighbors, $dn(\delta t)$, changes with time delay $\delta t$ in the flocking phase. This quantity remains constant in time for ideal solids but diverges for fluids. We find that the change of $dn(\delta t)$ for HaCaT cells is negligible compared to a typical cell size; its rate of growth is more than an order of magnitude slower than the one for MDCK cells. Similarly, Fig. 1E-H shows the probability of finding the same tagged cell at the location $(x,y)$ away from a reference HaCaT cell (E, F) or MDCK cell (G, H) at $\delta t$ = 0 and 170 min, respectively (see details in *Methods*), knowing that it started as a neighbor of the reference cell at $\delta t$ = 0. One finds that the probability pattern of MDCK cells slowly spreads with time delay (Fig. 1G and 1H) because of neighbor exchange events, while the spread is negligible for HaCaT cells (Fig. 1E and 1F, Movie S2). As a result, we conclude that HaCaT cells move coherently as a solid with negligible cell/cell rearrangements in the flocking phase. This mode of coherent motion is in stark contrast with most studies reported before for kidney and lung epithelial cells (such as MDCK and HBEC cells), where cells change their neighbors frequently and exhibit turbulent fluid-like dynamics (Movie S3-S5) (2). Since skin cells are more likely to be exposed to damages than kidney and lung cells, such a solid-like long-ranged coherent motion may be beneficial to functions such as wound healing. To assess the robustness of our findings, we extend our analysis to another type of skin cell, immortalized human keratinocytes (N/TERT1). Similar to what we find in HaCaT cells, N/TERT1 cells also move coherently with small rearrangements with respect to their neighbors, flocking like active solids (Fig. S1).

**Cell dynamics in different states.** Fig. 2 shows the dynamics of cells at different states I, II and III corresponding to Fig. 1A. In state I, cells start moving with a gradual increase of $v$ and $S_v$; in state II, cells move coherently as a solid and finally in state III, cells slow down and eventually jam. By performing immunostaining, we find that in state II, the cell-substrate adhesion protein α-6 integrin (red) is strongly polarized towards the back (relative to the flocking direction) while the actin cytoskeleton (cyan) is polarized towards the front. Such a global polarization is absent in states I and III (Fig. 2A-C). Correspondingly, the cells move slowly in random directions in states I and III, but exhibit a large-scale coherent movement in state II (Fig. 2D-F). Similar large-scale flocking motion as well as the polarization of α-6 integrin are also observed in N/TERT1 cells (Fig. S2 and Movie S6). It should be noted that the global direction of cell motion in state II is not deterministic and varies in each experiment, which is consistent with the absence of external cues guiding the dynamics, and a spontaneous breaking of rotational symmetry, as observed in, for example, Vicsek-like models (33).

We then apply traction force (44) and Bayesian inversion stress microscopy (45-47) to study how the intercellular forces change during the synchronization of cells. Fig. 2G-I shows that the traction forces between cells and the substrate as well as the stresses within cell collectives are strongly dependent on cell dynamics. The average magnitudes of the traction forces ($<|T|>$) and stresses ($<|P|>$) both slightly increase as the velocity order parameter $S_v$ increases, and drastically decrease before $S_v$ reaches its maximum (Fig. 1A and Fig. 2G). In addition, the traction forces and stresses show short-ranged polar and nematic order, respectively (Fig. S3). We calculate the spatial orientational correlation functions of the traction forces and stresses and derive the temporal dependence of their correlation lengths (see *Methods*). It is found that the polar correlation length of traction forces ($\lambda_T$) and the nematic correlation length of stresses ($\lambda_P$) exhibit similar profiles which are maximum at $t \approx$ 1800 min as $S_v$ increase, and then continuously decrease (Fig. 2G). This indicates that the intercellular forces are correlated to cell movements, and become more ordered as cells move coherently. However, collective motion in state II does not coincide with the maximal levels of traction force and stress; it occurs when the traction force and stress are decreasing and persists even at rather low values. Furthermore, we observe that when the cells move coherently in state II, the traction force and stress patterns propagate as waves in the direction of cell motion (Movie S7 and S8). The mechanical waves can be clearly distinguished in the kymograph shown in Fig. 2H and I. They propagate at a speed $c_{T,P} \sim 0.35 \pm 0.05$ μm min$^{-1}$, which is close to the mean speed of cells in the flocking phase, indicating that the traction force and stress patterns are advected by the cell flock.

**Physical properties of the long-ranged coherent motion.** Fig. 3A shows the cellular velocity order parameter, $S_v$, in different sizes of square regions of interest averaged over time as cells move coherently. We find that this order parameter satisfies $S_v - S_v^\infty \propto A^\beta$, where $A$ is the surface area of the square region of interest, $S_v^\infty = 0.75$ and $\beta = -0.39$ (Fig. 3A inset). Such an asymptotic behavior demonstrates that the order parameter converges to a finite positive value $S_v^\infty$ as $A$ goes to infinity, which is consistent with the definition of true long-range polar order. In addition, we compute $\delta v_i^\perp = v_i \times \sin\theta$, and $\delta v_i^\parallel = v_i \times \cos\theta - \langle v \rangle$ as the transverse and longitudinal components of the velocity fluctuations, where $\theta$ is the angle between the velocity of individual cells, $\mathbf{v}_i$, and the average velocity of cell collective $\langle \mathbf{v} \rangle$. We find that the correlation functions of the longitudinal velocity fluctuations along ($r_\parallel$) and perpendicular ($r_\perp$) to the global motion direction, $C_{\delta v\parallel}(r_\parallel)$ and $C_{\delta v\parallel}(r_\perp)$, are both short-ranged and decay to 0 within tens of micrometers (Fig. S4). In contrast, the correlation of the transverse velocity fluctuations is anisotropic (Fig. S5) and decays algebraically over up to two decades with exponents -0.52 and -0.34 along $r_\parallel$ and $r_\perp$, respectively as $r$ increases (Fig. 3B). Importantly, such an algebraic decay of the transverse velocity fluctuation correlation reveals the emergence of a massless Goldstone mode induced by the spontaneous breaking of the rotational invariance of velocity orientations (35, 48), which has so far only been reported in active fluids but not in active solids. Similar global polar velocity order and scale free velocity fluctuations are also observed in flocking N/TERT1 cells (Fig. S6). Another hallmark of Goldstone modes in active systems is provided by anomalous density fluctuations, where the standard deviation of the number of particles, $\sqrt{\langle \delta N^2 \rangle}$, in a given observation window grows faster than the square root of its mean, $\langle N \rangle$ ($\sqrt{\langle \delta N^2 \rangle} \propto \langle N \rangle^\alpha$ with $\alpha > 1/2$) (49-51); this results from the long range correlations of the velocity fluctuations, which together with self-propulsion lead to the breakdown of the central limit theorem. While so far these have been reported mostly in active fluid phases, we then wonder if such giant density fluctuations can also be observed in our active solid system. In Fig. 3C, we first subtract the average motion of cells with scale invariant feature transform method (SIFT) (52) and then characterize the mean and standard deviation of cell numbers averaged over time for different sizes of regions. It should be noticed that cell divisions are blocked here by mitomycin C. Giant density fluctuations with $\alpha \sim 0.8$ are observed, and consistently confirmed by the analysis of the structure factor (Fig. S7). We anticipate that due to the Goldstone mode of velocity fluctuations, which will induce increasingly large fluctuations of particle numbers and thus cell density at large scales, the epithelial monolayer can rupture during its collective motion. Indeed, such a phenomenon is observed in the flocking N/TERT1 cells (Fig. 3G, Fig. S8 and Movie S9).

Overall, our experimental findings reveal a collective mode of migration where a cell monolayer flocks as a solid with true long range polar order, and robust signatures of massless Goldstone mode such as scale-free velocity fluctuations and giant density fluctuations. While such features are generic of polar fluid phases, we show below that they are also consistent with a theoretical model of active polar solid.

**Theoretical model.** Based on our experimental findings, we propose a theoretical model to show that the observed solid flocking phases with massless rotational Goldstone modes and anomalous density fluctuations in cell monolayers are expected to be generic features of polar active solids, and thus shared by a broad range of active systems. The model is based on ref. (53) and provides a continuous coarse-grained description of general active, polar, cell monolayers assumed to be in a solid phase (see *Supplementary Materials* for details), typically because of sufficiently strong cell/cell junctions. It is phenomenological in nature and aims at providing a minimal physical mechanism that can account for the striking features of the observed solid flocking phases described above. To model the solid behavior of the epithelial sheet on the time scale of the observation, as reported in Fig. 1, together with the emergence of a polar flocking phase that spontaneously breaks rotational symmetry (see Fig. 1A, Fig. 2 and Fig. 3A), we introduce the two-dimensional displacement field $u(r,t)$ that accounts for in-plane elastic deformations of the monolayer and the polarization field $p(r,t)$ that accounts for in-plane cell polarity (see Fig. 3E and F). We model their coupled overdamped dynamics as follows:

$$\partial_t \bm{p} + v_0 \bm{p} \cdot \nabla \bm{p} = -\frac{1}{\gamma}\frac{\delta F}{\delta \bm{p}}, \qquad (1)$$

$$\partial_t \bm{u} + v_0 \bm{p} \cdot \nabla \bm{u} = v_0 \bm{p} - \frac{1}{\Gamma}\frac{\delta F}{\delta \bm{u}}, \qquad (2)$$

where $\gamma$ and $\Gamma$ are friction coefficients, $v_0$ the self-propulsion speed of the cells, and $F$ a free energy to be defined below. Note that here we use a Eulerian description that is standard for hydrodynamic theories but less frequent in elasticity theory: the implicit variable $r$ denotes the running coordinate of a given material point in the lab reference frame. To account for the phenomenology of the system in a minimal way, we keep the self-propulsion term as the only non-equilibrium one and neglect other possible active terms (53), as well as terms that would be higher order in gradients. Eq. (1) above describes the relaxation dynamics of the advected polarization field in response to variations of the free energy as would occur in classical liquid crystals, while Eq. (2) expresses the local force balance in the monolayer and comprises friction forces with the substrate, active self-propulsion, internal elastic forces, as well as advection. Following classical descriptions of polar elastic solids, the free-energy $F[\bm{p}, \bm{u}] = \int (f_u + f_p + f_{int}) d^2 r$ is taken to be the superposition of the elastic energy of the displacement field $f_u$, the Ginzburg-Landau energy of the polar field $f_p$, and the energy of interaction between the two fields $f_{int}$, which we write as

$$f_{int}(r, [\bm{u}, \bm{p}]) = \frac{w}{4}[\nabla \bm{p} + \nabla \bm{p}^T - \nabla \cdot \bm{p}\, \bm{I}] : [\nabla \bm{u} + \nabla \bm{u}^T - \nabla \cdot \bm{u}\, \bm{I}], \qquad (3)$$

where $\bm{I}$ is the identity matrix and " : " represents full tensor contraction. Importantly, for generic active polar solids there in principle exists an interaction term, compatible with the rotational symmetry of the combined fields, that is of lower order in gradients (see *Supplementary Materials*). We argue that in our cellular system this term must vanish as it would enslave the polarization to the displacement field, and thus make the orientational Goldstone mode massive. This would contradict the experimental observation of algebraic correlations of the transverse fluctuation of the speed $v \sim v_0 p$ (see Fig. 3B); we thus expect that within the time-and-length scales of the experiment, this coupling can indeed be neglected. Therefore, in Eq. (3), we retain only the lowest order terms that are invariant under *independent* rotations of a *uniform* $\bm{p}$ field and the $\bm{u}$ field; the finite time relaxation of the orientation field to displacement fluctuations is then avoided and the orientational Goldstone mode remains massless. In the context of cell monolayers, this choice implies qualitatively that the front back polarity of a cell is not elastically bound to the local cell environment (typically cell/cell junctions) and can freely rotate.

We now show that this minimal description indeed recapitulates their key features observed experimentally. We linearize the dynamics around the flocking solution $(p(r,t), u(r,t)) = (p_0, v_0 p_0 t)$ (see *Supplementary Materials* for details). We denote the parallel and transverse components of the fluctuations of each field by $(\delta u^{\parallel}, \delta u^{\perp})$ and $(\delta p, p_0 \theta)$, respectively. It is found that the amplitude fluctuations of the polarization, $\delta p$, relax exponentially fast and can be eliminated adiabatically. This allows for the explicit determination of the correlations of the spatial Fourier transform of $\delta u^{\parallel}$, $\delta u^{\perp}, \theta$ and thus the fluctuation of the transverse velocity $\delta v^{\perp}$, which are found to satisfy as $q \to 0$:

$$\langle \delta u_q^{\parallel} \delta u_{-q}^{\parallel} \rangle \sim \langle \theta_q \theta_{-q} \rangle \sim \langle \delta v_q^{\perp} \delta v_{-q}^{\perp} \rangle \sim \frac{1}{q^2} \; ; \quad \langle \delta u_q^{\perp} \delta u_{-q}^{\perp} \rangle \sim \frac{1}{q^4} \qquad (4)$$

First, such power-law dependence on $q$ shows the emergence of algebraic (or logarithmic) correlations in real space and thus of massless Goldstone modes, in agreement with Fig. 3B. Of note, we do not expect the value of the exponent 2 to be exact; its determination would require a full nonlinear analysis of the problem which is beyond the scope of this work. Second, this gives access upon Fourier inversion to the fluctuations of each field in a given observation window of area $A$:

$$\langle \delta u^{\parallel 2} \rangle \sim \ln A \; ; \; \langle \delta u^{\perp 2} \rangle \sim A. \qquad (5)$$

This divergence of fluctuations with the system size is in qualitative agreement with the results of Fig. 3D, even if the exact power-law dependence on A is not expected to be exactly determined by our linear analysis. Similarly, our model displays a static structure factor of density fluctuations that diverges as $1/q^2$ and, consequently, giant number fluctuations $\sqrt{\langle \delta N^2 \rangle} \propto \langle N \rangle$ that are qualitatively coherent with the results of Fig. S7 and Fig. 3C. Importantly, the strong divergence of the transverse displacement fluctuations implies that the positional order at large scales is washed away and the solid presumably either breaks or melts due to strains that diverge logarithmically. This is a direct consequence of the massless nature of Goldstone mode for the transverse polarity $\theta$; the invariance under rotation of a uniform polarity field independently of the displacement field, as we have assumed in defining $f_{int}$, allows $\theta$ to undergo arbitrary large fluctuations on sufficiently large scales which, in turn, lead to arbitrary large stresses in the medium through the self-propulsion forces (see Fig. 3E and F). This is consistent with the divergences reported in Fig. 3D, the small but non-zero increase of the average distance with initially neighboring cells shown on Fig. 1B, and, most importantly, with the rupture events in the epithelium reported in Fig. 3G, Fig. S8 and Movie S9. Finally, this shows that our minimal hydrodynamic theory of active polar solids correctly predicts the existence of flocking phases with distinct signatures of massless Goldstone modes such as algebraic fluctuations of the transverse velocity and giant density fluctuations, in agreement with our observations.

**Propagation of waves.** To further characterize the solid behavior of these epithelial cells, we reason that the propagation of mechanical waves in passive environments, including gas, liquids and solids is known to carry distinctive features of the underlying medium (54). In flocking active fluids, it was indeed predicated that flocks with long-ranged correlations could host two mixed sound-like modes, coupling density and polarization fluctuations, despite the inertialess dynamics (51). These waves were recently observed in active colloidal fluids (55) but remained to be elucidated in active solids. To determine the dispersion relation of sound modes in our epithelial solid, we analyze the space-time correlations of cell density in Fourier space, i.e. power spectra. As shown in Fig. S9 and Movie S10, we do not observe clear evidence for density waves. On the other hand, sound modes of the transverse components of velocity fluctuations (shear waves) are identified, which propagate in the opposite direction of the global motion of cells at a speed $c \approx$ -1.7 μm min$^{-1}$ (Fig. 4 and Movie S11). No waves of the longitudinal components of velocity fluctuations are observed (insets in Fig. 4B and C).

Interestingly, the dispersion relation can be obtained explicitly from the linearized dynamics of the model introduced above (see *Supplementary Materials*), and predicts the existence of three eigenmodes in the flocking phase. One of them is diffusive while the other two have propagating part. The existence of the latter, despite the non-inertial nature of the model, is allowed by the activity. Although the three fluctuating fields $\theta$, $\delta u^{\parallel}$ and $\delta u^{\perp}$ are entangled in these eigenmodes, a numerical analysis indicates that the diffusive mode is mainly supported by $\delta u^{\parallel}$ while the propagative modes are mainly carried by $\theta$ and $\delta u^{\perp}$. This is in accordance with the fact that shear waves are the only propagating waves to be experimentally observed, and is shown in *Supplementary Materials* to be exact in the limit of low compressibility. Furthermore, the damping coefficient of the propagating waves is found to be minimum in the longitudinal direction, which is coherent with the observation of shear waves that propagate mainly along the direction of the cell motion. Finally, the minimal model of polar elastic solid that we propose predicts the existence of shear waves for the transverse velocity that propagate along the flocking direction, which are indeed observed experimentally and could provide a further robust signature of solid flocking phases.

**Discussion and conclusion**
Most previous studies on the migration of cellular monolayers have been characterized within the framework of liquid flow-like systems that exhibited a turbulent-like collective motion with short-range velocity correlations (2, 6, 10). However, there is now growing evidence that epithelial cells can also exhibit large-scale polarized motion, for example during embryonic development of drosophila (56, 57). A pioneering work by Szabo et al. also showed that keratocytes (epithelial cells from fish scales) over a critical density could exhibit a large-scale coherent motion (29) and more recently Malinverno et al. demonstrated that epithelial cells, MCF-10A, experiencing an unjamming transition could move coherently in a relatively large spatial scale (27, 30). With the exception of a recent work by Lang et al. (40), such cellular flows have been so far usually associated to a fluid-like rheological description of the cell epithelium.

Here we demonstrate that two types of keratinocytes (HaCaT and N/TERT1) exhibit a seemingly comparable flocking behavior with long range order. It is important to note that although we utilize the same methodology and observe phenomena akin to those reported in reference (28), our study reveals the emergence of a spontaneous symmetry-breaking active solid phase which was not previously described. In particular, our analysis reveals that the cell monolayer moves as a polar elastic solid with minimal cell rearrangements, in striking contrast to the earlier descriptions of fluid flocking phases (33, 34). In addition, our experimental analysis uncovers that this solid flocking phase exhibits striking features of long-range polar order, with scale-free spatial correlations, anomalously large density fluctuations, and shear waves. On the basis of a general theory of active polar solids, we propose that these features arise from massless Goldstone modes, and provide robust signatures of solid flocking phases. Unlike in polar fluids where they are expected to be generic, these modes in solids require the decoupling of global polarity rotations from in-plane elastic deformations. In the context of cell epithelia, this implies that the front back polarity of individual cells is not slaved to the local cellular environment and can freely rotate, without involving cell/cell rearrangements; this prediction remains to be verified experimentally. Both theoretical predictions and experimental observations consistently show that in such polar active solid phases, elastic deformation fluctuations increase with system size, eventually leading to rupture (Fig. 3D-G and Fig. S8), which could have implications in the maintenance of tissue integrity as well as in the emergence of 3D shapes.

Altogether, our experimental and theoretical results suggest that such a solid-like long-ranged coherent motion broadly exists in many kinds of epithelial cells and can contribute to various fundamental biological functions such as in wound healing, embryonic development, and tissue morphogenesis. Also, it brings interesting questions about the different nature of epithelia coming from different tissues exhibiting fluid- or solid-like behaviors.

**Experimental Setup and Methods**
**Cell lines and culturing condition.** HaCaT cells (human keratinocyte cell line) (58) stably expressing histone2B-GFP were established by transfection of pRRlsinPGK-H2BGFP-WPRE (gift from Beverly Torok-Storb, Addgene plasmid # 91788) (58). Cells were grown in a culture dish. The diameter of the dish was 35 mm and was coated with 50 μg ml$^{-1}$ fibronectin. The cells were seeded at a density of ~4×10$^3$ cells per mm$^2$. Starvation of HaCaT cells and subsequent re-stimulation of cell cycle progression was performed as described in Ref (28). In brief, a confluent HaCaT cell layer was cultured in serum-free medium (Dulbecco's modified Eagle's medium, DMEM) for 2-3 days. Then the medium was replaced with DMEM supplemented with 15% fetal bovine serum (FBS). To characterize the density fluctuations, cell division was blocked by culturing cells in medium (DMEM + 15% FBS) containing 10 μg ml$^{-1}$ mitomycin C (Sigma Aldrich) for 1 hour, followed by washing with Dulbecco's phosphate-buffered saline (PBS) three times.

N/TERT1 keratinocytes (59) were a kind gift from the Niessen laboratory (CECAD Cologne, Germany). Cells are grown in growth medium (CnT-Prime, CellnTech, Switzerland). Cells were harvested and plated over night at high concentration at the center of the fluorodish, forming a large cluster which can spread in all directions. Differentiation was induced through changing the medium to CnT-Prime-3D, which contains high calcium and several growth factors. For immunostainings, cells were fixed after 24 hours of migration.

**Indirect immunostaining**

Fixation was carried out in 4% formaldehyde for 10 min at room temperature. Cells were permeabilized using 0.1% Triton X-100 in PBS for 5 min followed by 3 x 5 min washing in PBS. Samples were blocked with 1% BSA and 10 % FBS in PBS for 1 h at room temperature. The primary antibody against alpha-6 integrin (R&D systems, catalog no MAB13501) was diluted 1:100 in blocking solution and incubated for 2 h at room temperature or overnight at 4 ℃. The samples were washed 3 x 5 min with PBS and incubated with an anti-rat antibody conjugated to Alexa Fluor 647 diluted 1:200 in blocking solution for 2 h at room temperature. Subsequently, samples were washed 3 x 5 min with PBS. The actin cytoskeleton was visualized using Phalloidin-Alexa Fluor 568 (catalog no A12380, Life Technologies) diluted 1:200 in PBS and the nuclei using Hoechst 33342 (catalog no 62249, Thermo Fisher) diluted 1:2000 in PBS for 45 min at room temperature. Image acquisition was carried out on a Nikon CSU-W1 spinning disc microscope or a Zeiss LSM 980 confocal scanning microscope equipped with an Airyscan 2 module.

**Live cell imaging and data analysis**. Samples were observed through a 10× objective on a BioStation IM-Q (Nikon, Tokyo, Japan) at 37 °C and 5% $CO_2$ with humidification. Only the center regions of the sample are characterized through the microscope. Images were taken every 10 min or 15 min for more than 40 hours. The movies were then analyzed through Cellpose2.0 (60), ImageJ and MATLAB. Basically, nuclei were first segmented by Cellpose, which were then tracked using the Trackmate plugin of ImageJ. The tracking results were then analyzed through MATLAB.

The average distance between cell $i$ and its neighbors shown in Fig. 1B is defined as $dn(t + \delta t) = \langle\langle \sum_i^{N(t)} \sum_j^{n_i(t)} \|\mathbf{r}_i(t + \delta t) - \mathbf{r}_j(t + \delta t)\| \rangle_{r_{ij}}\rangle_t$, where $\delta t$ is time delay, $N(t)$ is the total number of cells in the field of view at time $t$, $n_i(t)$ is the number of voronoi neighbors of cell $i$ at $t$, $r_i(t+\delta t)$ and $r_j(t+\delta t)$ are the coordinates of cell $i$ and cell $j$ at time $t+\delta t$. The probability density distribution shown in Fig. 1E-H characterize the probability of finding the same tagged cell at the location ($x,y$) away from a reference cell at different time delays, knowing that it started as a neighbor of the reference cell at $\delta t = 0$, is defined as $P(r, \delta t) = \langle\langle \sum_i^{N(t)} \sum_j^{n_i(t)} \Delta\left(r_{ij}(t + \delta t) - r_{ij}(t)\right)\rangle_{r_{ij}}\rangle_t$, where $\delta t$ is time delay, $N(t)$ is the total number of cells in the field of view at $t$, $n_i(t)$ is the number of neighbors of cell $i$ at $t$, $r_{ij}(t)$ is the distance between cell $i$ and cell $j$ at time $t$. The spatial orientational correlation functions of traction forces and stresses are defined as $C(r) = \langle \varphi(0)\varphi(r) \rangle$, where $\varphi$ represents the orientational vector of traction forces or stresses. The corresponding correlation length, $\lambda$, are calculated by fitting the correlation function with exponentially decaying function $y = a*\exp(-r/\lambda)$, where $a$ is a constant. The correlation function of velocity fluctuations is defined as $C_{\delta v \perp, \parallel}(r) = \langle\langle \delta v_i^{\perp,\parallel}(t) \delta v_j^{\perp,\parallel}(t) \rangle_{\mathbf{r}_i - \mathbf{r}_j}\rangle_t$, where $\delta v_i^{\perp,\parallel}(t)$ are the transverse and/or longitudinal components of the velocity fluctuations. The one-point correlator of cell displacements in the transverse ($\delta u^\perp$, red circles) and longitudinal ($\delta u^\parallel$, black squares) directions of the global motion is defined as the variance of the components of the mean displacement of cells perpendicular to and along the global motion direction of cells in a specific region of surface area $A$, i.e., $C_{\delta u^{\parallel,\perp}} = \langle \left(\delta u_i^{\parallel,\perp}\right)^2 \rangle_A - \langle \delta u_i^{\parallel,\perp} \rangle_A^2$, where $u_i^\parallel = (r_i(t) - \langle r_i \rangle_t) \times \cos\theta_i(t)$, $u_i^\perp = (r_i(t) - \langle r_i \rangle_t) \times \sin\theta_i(t)$, $\theta_i(t)$ is the angle between the velocity of cell $i$ and the global velocity of cell collective at time $t$. The spatial Fourier transform of the velocity fluctuation field is defined as $V_q(t) = \sum_i \delta v_i(t) e^{iq[x_i(t)\cos\theta + y_i(t)\sin\theta]}$, where $x_i(t)$ and $y_i(t)$ are the instantaneous coordinates of cells, $q(\cos\theta, \sin\theta)$ is the wave vector making an angle $\theta$ with respect to the global order. The power spectra $|V_{q,\omega}|^2$ are obtained by performing time Fourier transformations of the two-time autocorrelations of the velocity fields $\langle V_q(t) V_q^*(t + \delta t)\rangle_t$. Similarly, The spatial Fourier transform of the density field is defined as $\rho_q(t) = \sum_i e^{iq[x_i(t)\cos\theta + y_i(t)\sin\theta]}$. The power spectra $|\rho_{q,\omega}|^2$ are obtained by performing time Fourier transformations of the two-time autocorrelations of the density fields $\langle \rho_q(t) \rho_q^*(t + \delta t)\rangle_t$. The static structure factor is defined as $S(q) = \langle \sum_{j,k} e^{iq(r_j - r_k)} \rangle / N$, where $N$ is the total number of cells in the field of view, $\boldsymbol{q}$ is wave vector.

**Author Contributions:**
Y. S. conceived and carried out the experimental investigations, analyzed the experimental results and wrote the draft of the manuscript. J. O. B. developed the theoretical model. A. S. carried out the experiments of human N/TERT1 keratinocytes. B. L. supervised the project. All authors contributed to discussions and writing the manuscript.

**Acknowledgements**
We would like to thank Joseph d'Alessandro, Lucas Anger, Marc-Antoine Fardin, Carien Niessen, Matthias Rübsam and the members of the "Cell adhesion and Mechanics" team for helpful discussions. This work was supported by the European Research Council (Grant No. Adv-101019835 to 488 BL and Synergy grant Shapincellfate to RV), LABEX Who Am I? (ANR-11-LABX-0071 to BL and RMM), the Ligue Contre le Cancer 489 (Equipe labellisée 2019 to RMM), the Agence Nationale de la Recherche ("STRATEPI" DFG-ANR-22-CE92-0048 to RMM) and PSAM to AM, the CNRS through 80 Prime program (to AS, BL) and Human Frontier Science Program (grant number LT0007/2023-C) (To YS). We acknowledge the ImagoSeine core facility of the IJM, a member of IBiSA and France-478 BioImaging (ANR-10-INBS-04) infrastructures.

**Data availability**
The data that support the findings of the study are available from the corresponding authors upon reasonable request.

**Competing interests**
The authors declare no competing interests.

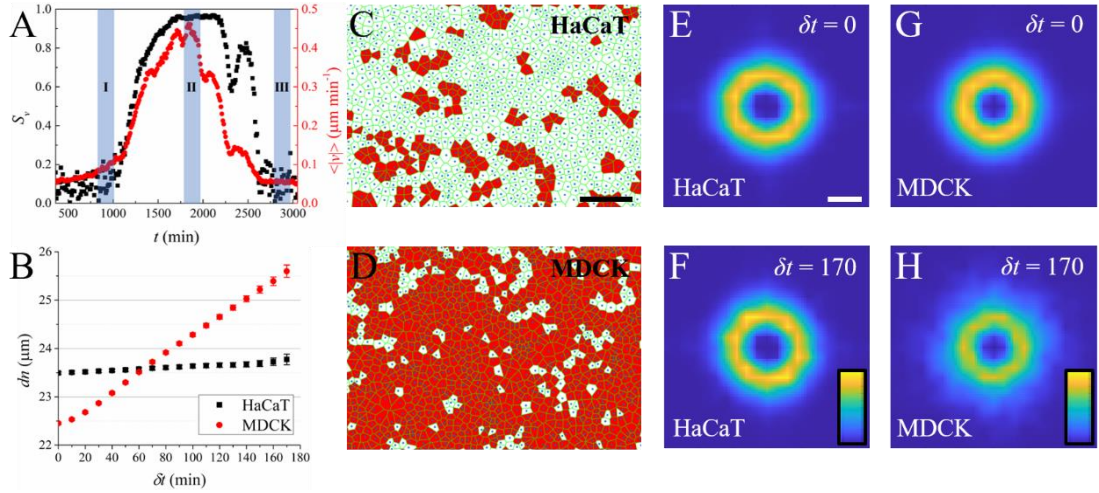

**Fig. 1** (A) The temporal evolution of the velocity order parameter ($S_v$, black squares) and the magnitude of average speed ($<|v|>$, red circles) of HaCaT cells. The measurement started at 360 min after serum stimulation. The three shadowed regions, I, II, and III, represent three dynamical regimes described in Fig.2. (B) The mean separation distance ($dn$) between cells (black: HaCaT cells; red: MDCK cells) and their initial neighbors as a function of time delay ($\delta t$). The Voronoi diagrams of HaCaT (C) and MDCK (D) cells, respectively. The cells colored by red represent the floppy cells (defined in the main text). Scale bar 200 μm. (E)-(H) The probability per unit area of finding the original neighbors at the location ($x,y$) away from the reference HaCaT cell (E, F) and MDCK cell (G, H) at different time delays (E, G: $\delta t = 0$ min, F, H: $\delta t = 170$ min). Scale bar 20 μm. The color bar changes linearly from 0 (dark blue) to 0.2 (light yellow).

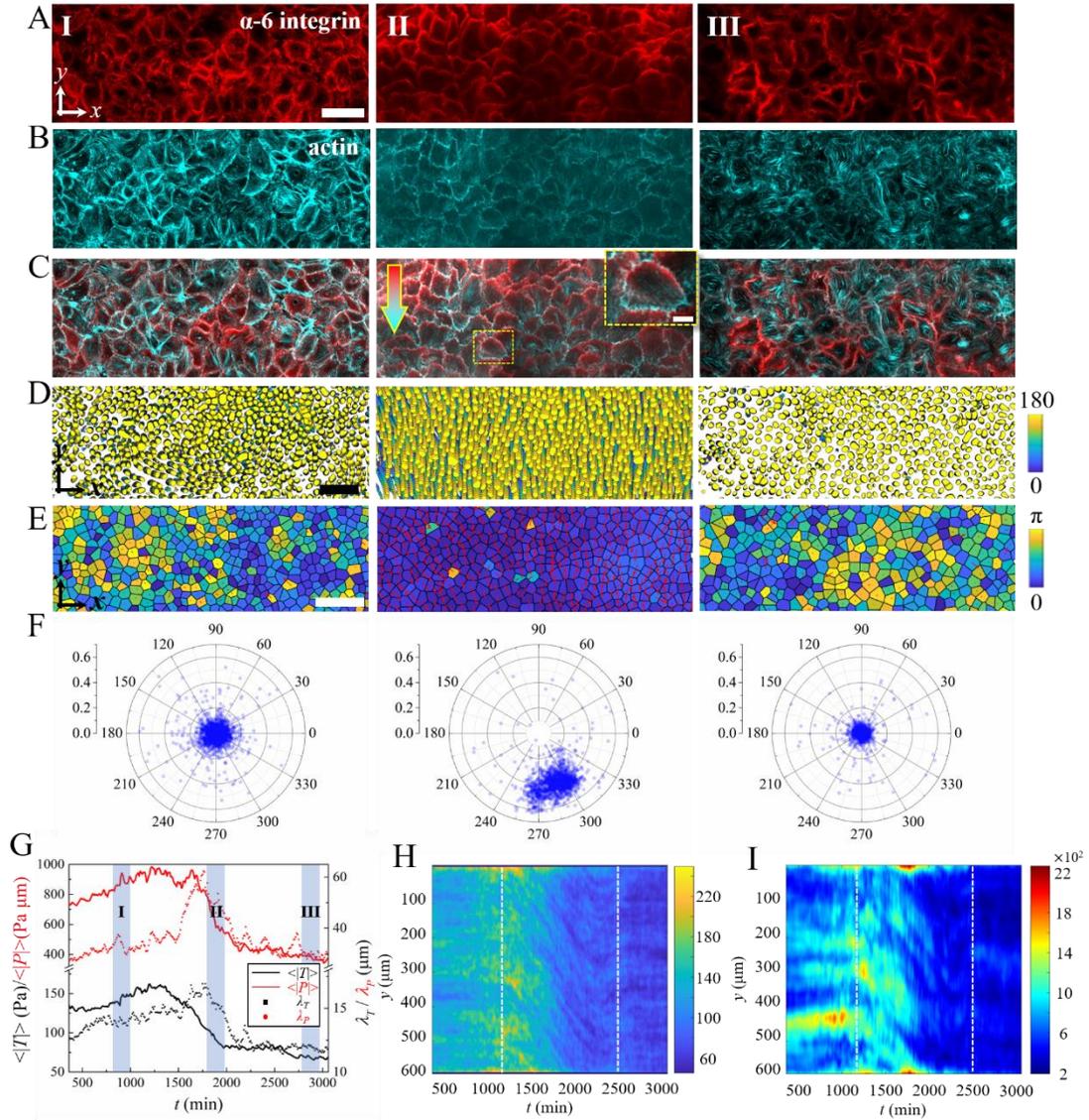

**Fig. 2** Confocal images of α-6 integrin (A), actin (B), α-6 integrin (red) and actin (cyan) merged (C) immunostaining of HaCaT cells fixed in states I, II and III. The arrow in (C) indicates the collective motion direction of cells. Scale bar 50 μm. The inset in (C) shows the zoomed-in image of a typical cell. Scale bar 10 μm. (D) Trajectories of cells in states I, II and III. The nuclei are colored with time corresponding to the color bar which changes linearly from 0 min to 180 min. Scale bar 100 μm. (E) Snapshots of cells at different moments in states I, II and III. The contours of cells are plotted as the voronoi diagrams which are colored according to the angle between the velocity of the cell and the average velocity of cells in the field of view. Scale bars 100 μm. (F) The polar distribution of the velocity of the cells shown in (E). (G) Temporal evolution of the magnitudes (solid lines) and correlation lengths (dots) of traction forces (black) and stresses (red). (H) and (I) Kymographs of traction forces and isotropic stresses, respectively.

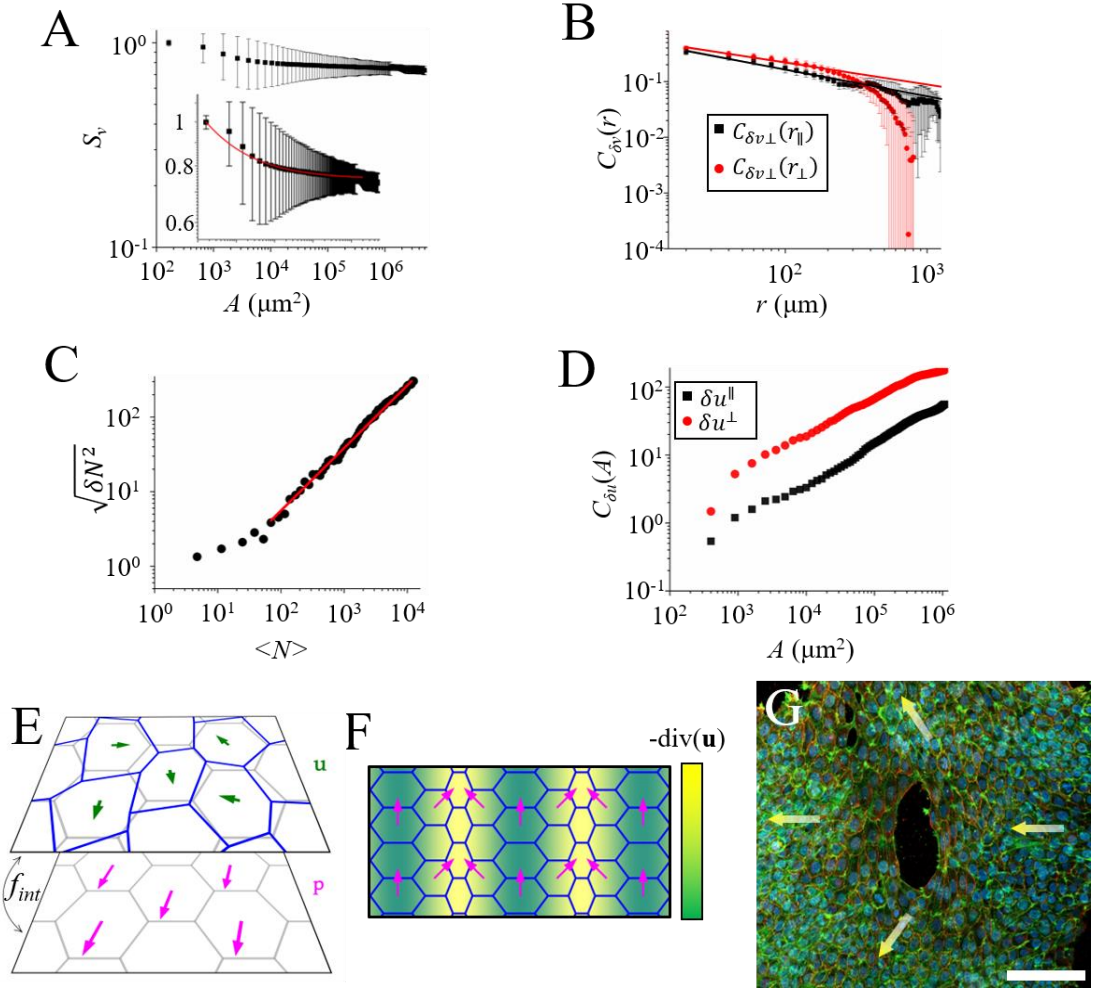

**Fig. 3** (A) The dependence of the velocity order parameter, $S_v$, on the surface area of the region of interest, $A$. The inset shows a magnified view of the data. The red line shows a nonlinear fit $S_v = S_v^\infty + kA^\beta$ with $k = 1.8$, $\beta = -0.39$ and $S_v^\infty = 0.75$. (B) Log-log plot of the correlation function of $\delta v_i^\perp$ in the transverse ($r_\perp$, red circles) and longitudinal ($r_\parallel$, black squares) directions of the global motion. The black and red lines are power law fits with exponents of -0.52 and -0.34, respectively. (C) Log-log plot of the standard deviation of cell number $\sqrt{\langle \delta N^2 \rangle}$ versus the mean cell number $<N>$ during the coherent motion. The red line is the power law fit of the experimental data with an exponent of 0.84. (D) Log-log plot of the dependence of the one-point correlator of cell displacements in the transverse ($\delta u^\perp$, red circles) and longitudinal ($\delta u^\parallel$, black squares) directions of the global motion on the surface area of regions of interest (see *Methods*). (E) A schematic picture of the model. The gray and blue lattices on the upper plane respectively represent the undeformed and deformed configurations of the elastic solid, the difference between the two being quantified by the displacement field $u(r, t)$ (green arrows). On the lower plane, the magenta arrows stand for the polarization field $p(r, t)$. The curved black arrow between the two planes symbolizes the interaction free-energy density $f_{int}(r, [u, p])$ that couples the two fields. The higher-order interaction energy that we choose here turns the angular rotational part $\theta(r, t)$ of the polarization field into a massless Goldstone mode. As depicted in the (F), the $\theta$ field can then undergo spin waves of arbitrary small energetic cost, which could, in turn, induce large deformations of the tissue, hence destabilizing the solid. (G) Confocal image of coherently moving human N/TERT1

keratinocytes. Green: actin. Blue: cell nucleus. Red: α-6 integrin. The yellow arrows indicate the local velocity direction of cells. One can clearly see the formation of a hole in the center. Scale bar 100 μm.

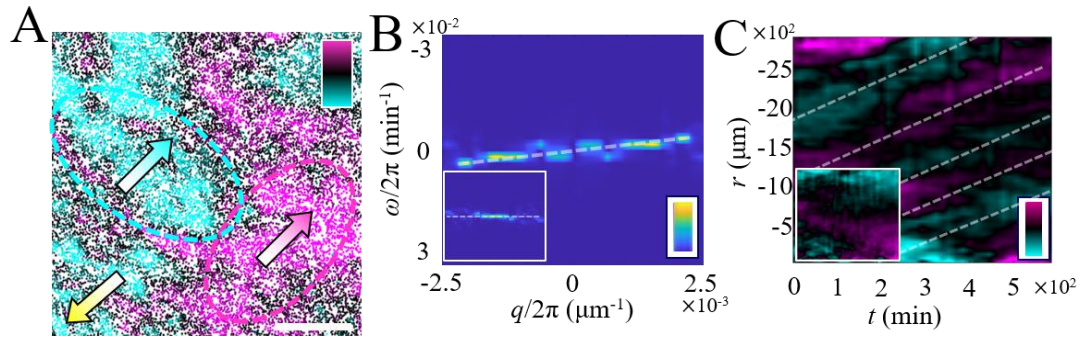

**Fig. 4** (A) A snapshot of the propagation of the transverse component of velocity fluctuation waves. Magenta and cyan colors represent the positive and negative transverse components of velocity fluctuations, respectively. The yellow arrow represents the direction of global motion of cells. The magenta and cyan arrows represent the propagation direction of velocity fluctuation waves. The color bar scales linearly from -0.3 to 0.3 μm min$^{-1}$. Scale bar 500 μm. (B) Power spectrum of $\delta v_i^{\perp}$ and $\delta v_i^{\parallel}$ (inset). The experimental data is first aligned with SIFT to subtract the contribution of the global motion of cells. The color bar scales linearly from 0 to 1. (C) Kymographs of $\delta v_i^{\perp}$ and $\delta v_i^{\parallel}$ (inset). The experimental data is first aligned with SIFT. The $r$-axis represents the axis along the global motion direction. The color bar scales linearly from -0.2 to 0.2 μm min$^{-1}$. The white dashed lines in (B) and (C) are guides to eyes.

# Supplementary Materials

## Scale-free flocking and giant fluctuations in epithelial active solids

Yuan Shen[1]*, Jérémy O'Byrne[2], Andreas Schoenit[1], Ananyo Maitra[2,3], Rene-Marc Mege[1], Raphael Voituriez[2]*, and Benoit Ladoux[1,4,5]*

[1] Universite Paris Cite, CNRS, Institut Jacques Monod, F-75013 Paris, France.
[2] Laboratoire Jean Perrin, CNRS, Sorbonne Université, Paris, France
[3] LPTM, CNRS/CY Cergy Paris Université, F-95032 Cergy-Pontoise cedex, France
[4] Department of Physics, Friedrich-Alexander Universität Erlangen-Nürnberg, Erlangen, Germany
[5] Max-Planck-Zentrum für Physik und Medizin, Erlangen, Germany

* Correspondence: yuan.shen@ijm.fr, raphael.voituriez@sorbonne-universite.fr and benoit.ladoux@fau.de

**This file includes the following sections:**

1. Supplementary Figures
2. Legends of Supplementary Movies
3. Details about theoretical model

# Supplementary Figures

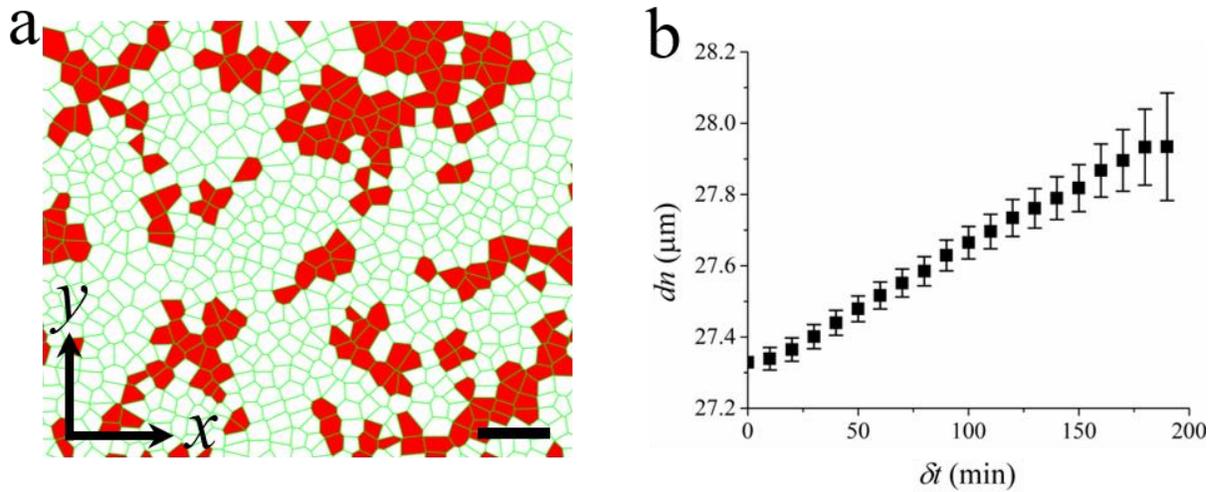

**Fig. S1.** The solid-like behavior of human N/TERT1 keratinocytes during coherent motion. (a) The voronoi diagrams of keratinocytes. The cells colored by red represent the floppy cells (defined in the main text). Scale bar 100 μm. (b) The mean separation distance ($dn$) between cells and their initial neighbors as a function of time delay ($\delta t$).

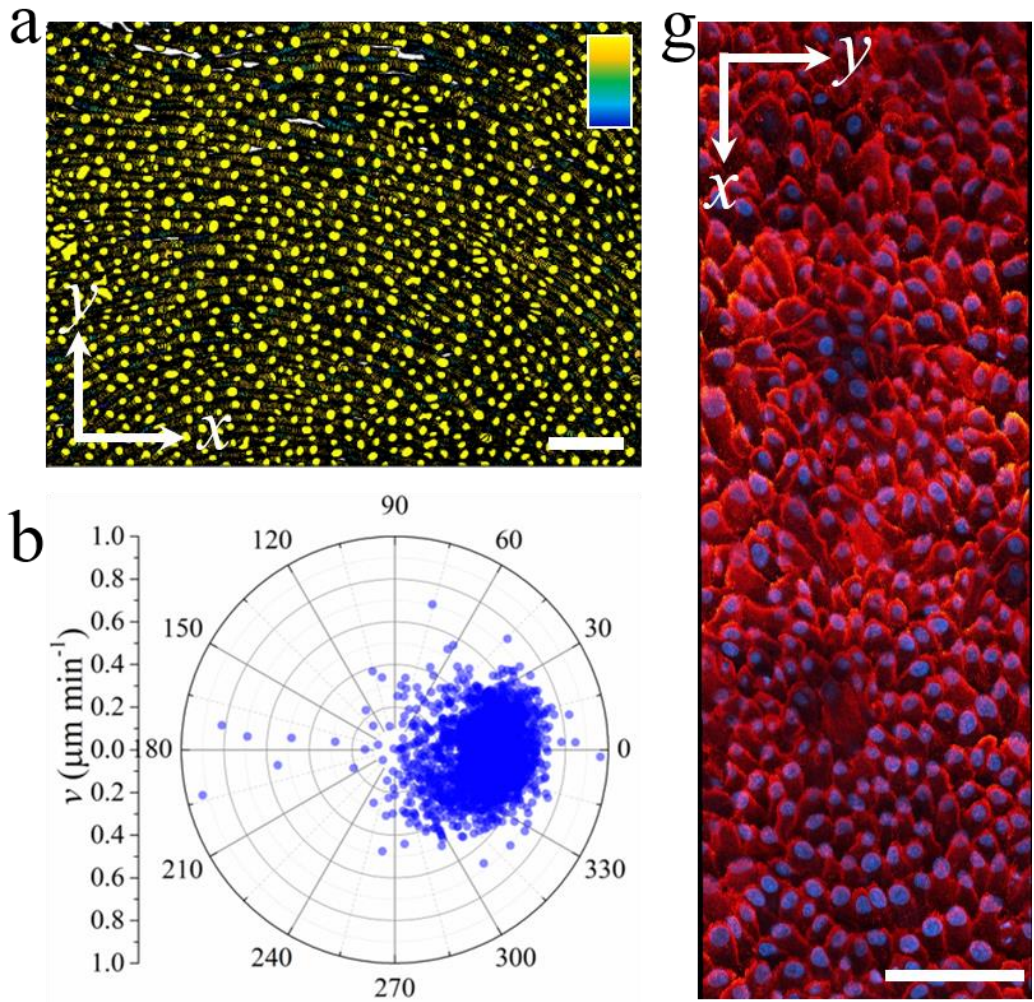

**Fig. S2.** The flocking motion of human N/TERT1 keratinocytes. (a) The trajectory of the nuclei of cells. The color bar scales linearly from 0 min to 200 min. Scale bar 100 μm. (b) The polar distribution of the velocity of the cells. (c) Confocal image of the immunostaining of keratinocytes. Red: α-6 integrin. Blue: cell nucleus. Scale bar 100 μm.

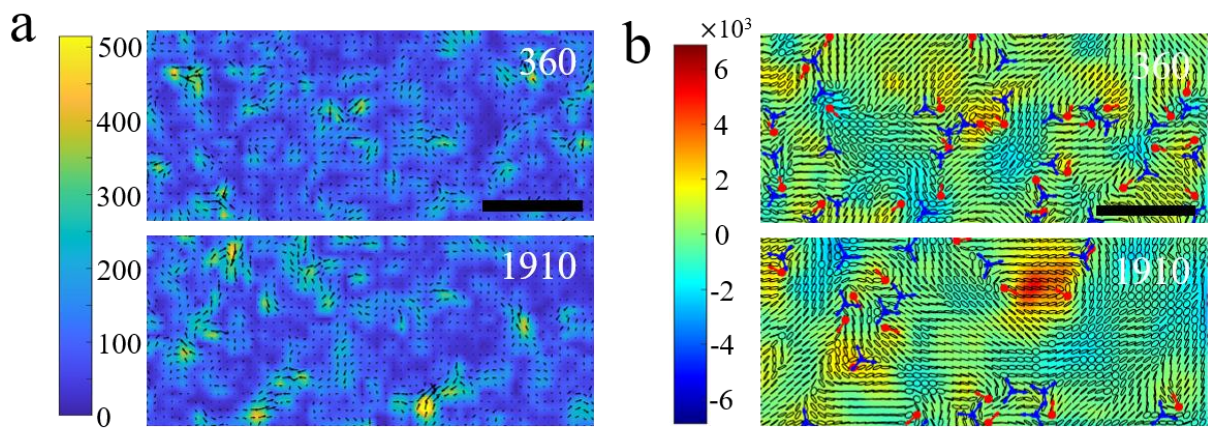

**Fig. S3.** Snapshots of traction force field (a) and stress field (b) of HaCaT cells at different moments ($t$ = 360 min and 1910 min, respectively). Scale bars 128 μm. The traction forces show local polar order whose magnitude and orientation are represented by the color map and black arrows, respectively. The stresses show local nematic order whose magnitude and orientation are represented by the color map and ellipses, respectively.

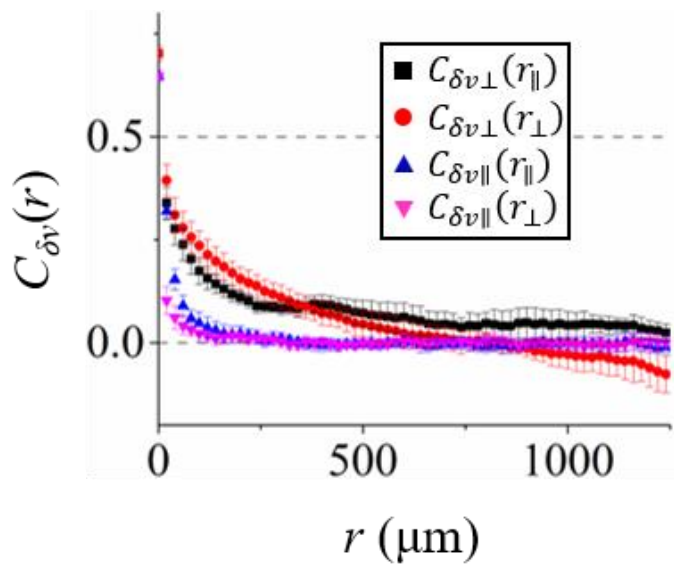

**Fig. S4.** The correlation functions of $\delta v_i^\perp$ (red circles and black squares) and $\delta v_i^\parallel$ (blue and pink triangles) in the transverse ($r_\parallel$) and longitudinal directions ($r_\perp$) of the global motion, respectively.

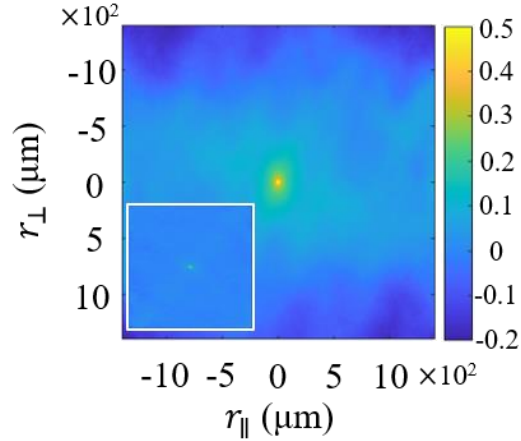

**Fig. S5.** The two-dimensional correlation function of the transverse component $\delta v_i^\perp$ and the longitudinal component $\delta v_i^\parallel$ (inset) of velocity fluctuations.

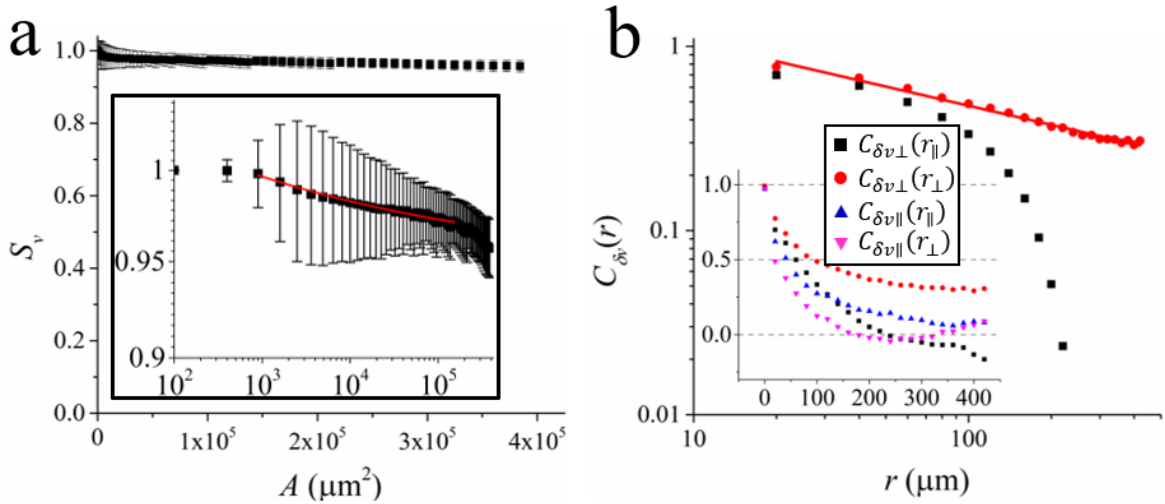

**Fig. S6.** (a) The dependence of the velocity order parameter, $S_v$, on the surface area of the region of interest, $A$. The inset shows the log-log plot of the data. The red line shows a nonlinear fit $S_v = S_v^\infty + kA^\beta$ with $k = 0.13$, $\beta = -0.14$ and $S_v^\infty = 0.95$. (b) Log-log plot of the correlation function of $\delta v_i^\perp$ in the transverse ($r_\perp$, red circles) and longitudinal ($r_\parallel$, black squares) directions of the global motion. The red line is a power law fitting with an exponent of -0.35. Inset: the correlation functions of $\delta v_i^\perp$ (red circles and black squares) and $\delta v_i^\parallel$ (blue and pink triangles) in the transverse and longitudinal directions of the global motion, respectively.

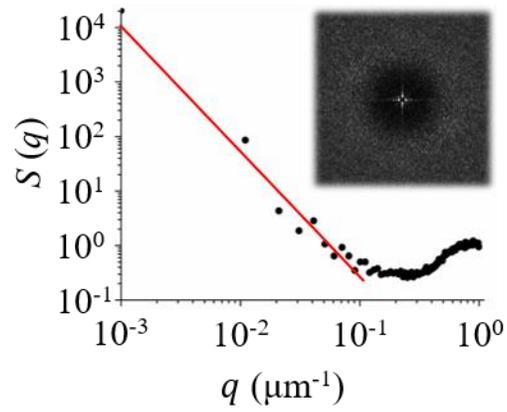

**Fig. S7.** Log-log plot of the static structure factor of cell nuclei. The red line is the power law fit of the experimental data with an exponent equal to -2.3. The inset shows the corresponding two-dimensional pattern.

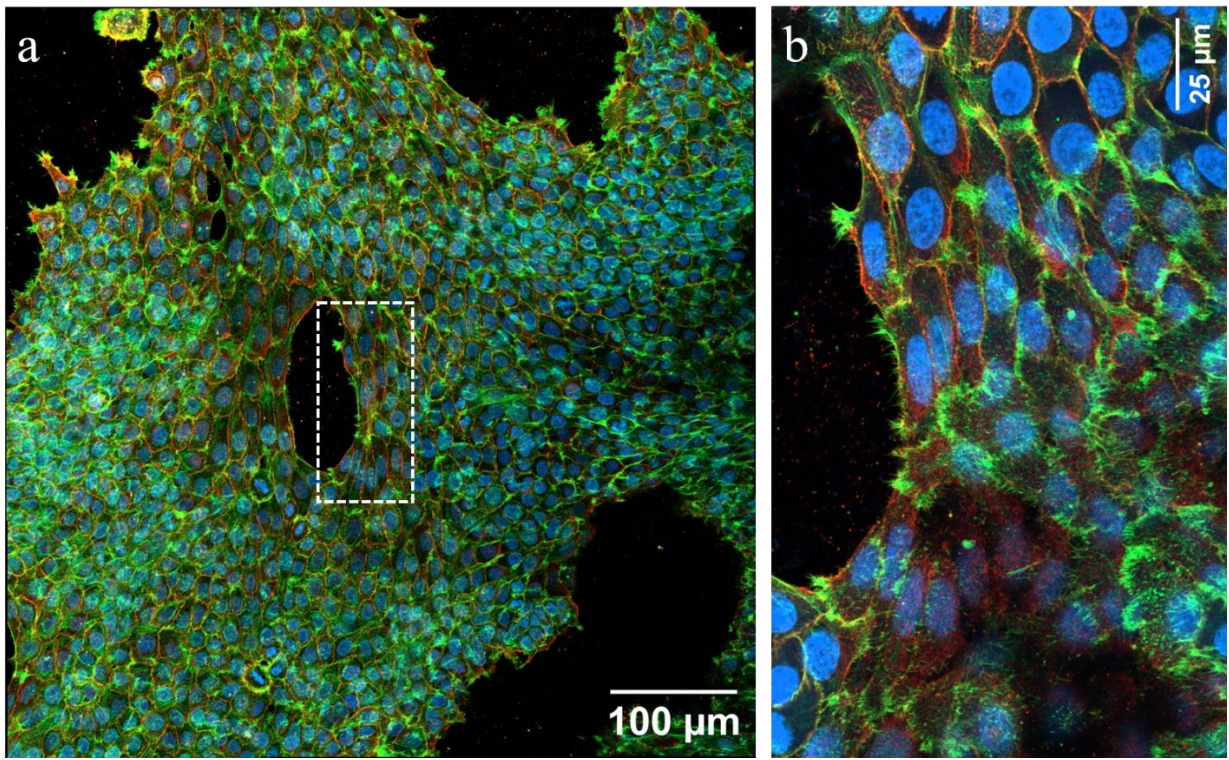

**Fig. S8.** (a) Confocal image of coherently moving human N/TERT1 keratinocytes. Green: actin. Blue: cell nucleus. Red: α-6 integrin. (b) The magnified view of the region inside the white dashed square in (a).

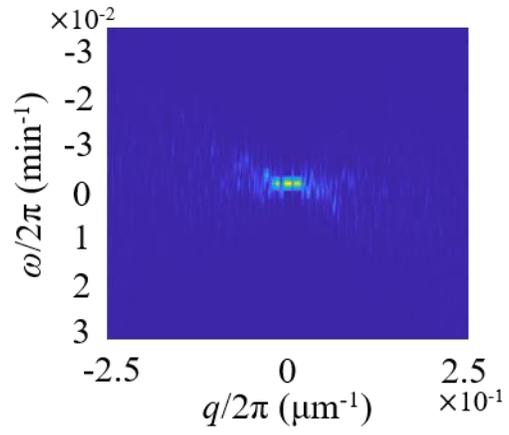

**Fig. S9.** Power spectrum of density fluctuations of HaCaT cells during the flocking phase.

## Supplementary Movies

**Movie S1.** The collective motion of HaCaT cells corresponding to Fig. 1A.

**Movie S2.** The temporal evolution of the probability density distribution pattern shown in Fig. 1 E-H.

**Movie S3.** The collective motion of HaCaT cells in stage II and the collective motion of MDCK cells.

**Movie S4.** The collective motion of HaCaT cells in stage II and the collective motion of MDCK cells. The global motion of cells is subtracted by the scale invariant feature transform (SIFT) method.

**Movie S5.** The voronoi diagrams of the HaCaT cells in stage II and the MDCK cells and their corresponding dynamical trajectories. The color bar indicates the frame number which increases from 0 to 17. Frame rate 10 min per frame. The global motion of cells is subtracted by the scale invariant feature transform (SIFT) method.

**Movie S6.** The collective motion of human N/TERT1 keratinocytes.

**Movie S7.** The time-stack color maps of the traction force field of HaCaT cells.

**Movie S8.** The time-stack color maps of the stress field of HaCaT cells.

**Movie S9.** Formation of holes during the collective motion of human N/TERT1 keratinocytes.

**Movie S10.** Collective motion of HaCaT cells. Cells are colored with density fluctuations. The left one is aligned with SIFT technique and the right one is not aligned.

**Movie S11.** Collective motion of HaCaT cells. Cells are colored with the transverse component of velocity fluctuations. The motion of cells is aligned with SIFT technique.

# Supplementary Material : Model section
# Scale-free flocking and giant fluctuations in epithelial active solids

Shen et al.

## I. INTRODUCTION

We model the epithelial monolayer as a two-dimensional active polar solid, where in plane elastic deformations are accounted for by a displacement field $\mathbf{u}(\mathbf{r}, t)$, while the polarity of the cells is represented by a polarization field $\mathbf{p}(\mathbf{r}, t)$, their velocity by $\mathbf{v}(\mathbf{r}, t)$ and density by $\rho(\mathbf{r}, t)$. Here $\mathbf{r}$ is the running spatial coordinate of a material point of the epithelium in a fixed frame of reference and we will write the elasticity and the dynamical equations of motion in Eulerian coordinates.

The solid is assumed to *spontaneously* break rotation symmetry and travel in the direction of the polarization. Retaining only the lowest order terms in gradients, the deterministic part of the dynamics can be written

$$\partial_t \rho + \nabla \cdot (\rho \mathbf{v}) = 0 , \tag{1}$$

$$-\Gamma \mathbf{v} + v_0 \Gamma \mathbf{p} - \frac{\delta \mathcal{F}}{\delta \mathbf{u}} = 0 , \tag{2}$$

$$\partial_t \mathbf{p} + \mathbf{v} \cdot \nabla \mathbf{p} = -v_p \mathbf{p} \cdot \nabla \mathbf{p} - \frac{1}{\gamma} \frac{\delta \mathcal{F}}{\delta \mathbf{p}} , \tag{3}$$

$$\partial_t \mathbf{u} + \mathbf{v} \cdot \nabla \mathbf{u} = \mathbf{v} . \tag{4}$$

Equation (1) is the conservation of mass with mass density $\rho$ ans velocity $\mathbf{v}$. Equation (2) corresponds to force balance equation, where inertia and viscosity have been neglected, with $\Gamma$ being the substrate-tissue friction coefficient, and $v_0$ the self-propulsion speed. In equation (3), the polarization relaxes at a rate $\gamma^{-1}$ to minimize the "free-energy" $\mathcal{F}$ and is advected by both the velocity field and by itself at speed $v_p$. Neither $v_p$ nor $v_0$ exists in passive systems and they are independent effects of activity; in the main text, we primarily discussed the effect of $v_0$ in the spirit of minimalism. We will primarily focus on this case even here, only briefly commenting on the effect of $v_p$ on the mode structure. Equation (4) relates the displacement field to the velocity, assuming that $\mathbf{u}$, which is associated to the solid structure, moves only with the mass motion and not *relative* to it [4]. This further implies that the fluctuations of the density field is slaved to the displacement field; upon linearizing around a homogeneous profile $\rho = \rho_0 + \delta\rho$, this reads

$$\delta\rho = -\rho_0 \nabla \cdot \mathbf{u} . \tag{5}$$

The "free-energy functional" $\mathcal{F}$ reads $\mathcal{F}[\mathbf{p}, \mathbf{u}] = \int d\mathbf{r} \, [f_\mathbf{p} + f_\mathbf{u} + f_{\text{int1}} + f_{\text{int2}}]$, with the densities

$$f_\mathbf{p}(\mathbf{r}, [\mathbf{p}]) = \frac{a}{2}|\mathbf{p}|^2 + \frac{b}{4}|\mathbf{p}|^4 + \frac{\kappa}{2}|\nabla \mathbf{p}|^2 , \tag{6}$$

$$f_\mathbf{u}(\mathbf{r}, [\mathbf{u}]) = \frac{B}{2}\left|\frac{\nabla \mathbf{u} + \nabla \mathbf{u}^\top}{2} - \frac{\nabla \cdot \mathbf{u}}{2}\mathbf{I}\right|^2 + \frac{B'}{2}(\nabla \cdot \mathbf{u})^2 , \tag{7}$$

$$f_{\text{int1}}(\mathbf{r}, [\mathbf{p}, \mathbf{u}]) = w \left[\frac{\nabla \mathbf{p} + \nabla \mathbf{p}^\top}{2} - \frac{\nabla \cdot \mathbf{p}}{2}\mathbf{I}\right] : \left[\frac{\nabla \mathbf{u} + \nabla \mathbf{u}^\top}{2} - \frac{\nabla \cdot \mathbf{u}}{2}\mathbf{I}\right] + w'(\nabla \cdot \mathbf{p})(\nabla \cdot \mathbf{u}) , \tag{8}$$

$$f_{\text{int2}}(\mathbf{r}, [\mathbf{p}, \mathbf{u}]) = c(\mathbf{p} \otimes \mathbf{p}) : \left[\frac{\nabla \mathbf{u} + \nabla \mathbf{u}^\top}{2} - \frac{\nabla \cdot \mathbf{u}}{2}\mathbf{I}\right] . \tag{9}$$

In eqs. (6)-(9), $\nabla$ is the gradient operator, $\nabla \cdot$ is the divergence operator, $|\ldots|$ is the norm of the appropriate tensor space[1], ':' stands for the full tensor contraction, and $(\ldots)^\top$ is the matrix transposition.

When polar symmetry is spontaneously broken in a previously isotropic solid, it becomes anisotropic [3, 5, 6]. Importantly, the polarisation fluctuations transverse to the ordering direction, which are the Nambu-Goldstone modes corresponding to broken rotation symmetry relax within a finite time to a value determined by displacement fluctuations via an Anderson-Higgs-like mechanism [1, 3]. The full physics of such a state in an active system is discussed in detail in [4].

---

[1] Each norm is the square-root of the sum of the square of all components in an orthonormal basis.

However, the time to realise the hydrodynamic behaviour of active polar solids is controlled by the free energy coupling parametrised by $c$ and is $\propto 1/c$. If $c$ is very small (compared to the other free energetic and active parameters), this timescale is very large and there is a long transient during which the the transverse fluctuations of the polarisation remain massless and hydrodynamic. Our experimental system seems to exist in this regime as the measurements of transverse polarisation fluctuations demonstrate that these are long-ranged. Therefore, here we discuss the physics of a "free-spin" polar solid in which the anglular fluctuations are not gapped, with the understanding that at very long times, the physics described here will crossover to the one described in [4]. Another way of anticipating that the polarization becomes "free to rotate" upon neglecting $f_{\text{int}2}$ is to note that the remaining coupling free-energy density, $f_{\text{int}1}$, when restricted to uniform polarizations, is invariant under *independent* rotations of the polarization and displacement fields.

Ignoring $c$ and solving for the velocity field from (2), the eqs. (3) & (4) explicitly read

$$\gamma \left[ \partial_t \mathbf{p} + (v_0 + v_p) \mathbf{p} \cdot \nabla \mathbf{p} \right] = -(a + b|\mathbf{p}|^2) \mathbf{p} + \kappa \Delta \mathbf{p} + \frac{w}{2} \Delta \mathbf{u} + w' \nabla (\nabla \cdot \mathbf{u}) \ , \tag{10}$$

$$\Gamma \left[ \partial_t \mathbf{u} + v_0 \mathbf{p} \cdot \nabla \mathbf{u} \right] = v_0 \Gamma \mathbf{p} + \frac{B}{2} \Delta \mathbf{u} + B' \nabla (\nabla \cdot \mathbf{u}) + \frac{w}{2} \Delta \mathbf{p} + w' \nabla (\nabla \cdot \mathbf{p}) \ , \tag{11}$$

to $\mathcal{O}(\nabla^2)$, where $\Delta$ is the Laplacian. Assuming in addition $v_p = 0$ yields Eqiuations (1) and (2) of the main text.

## II. LINEARIZED DYNAMICS

We now examine the linearised dynamics implied by our free-spin polar solid [2] in the polarised state realised when $a < 0$. A solution of eqs. (10) & (11) is given by

$$(\mathbf{p}(\mathbf{r}, t), \mathbf{u}(\mathbf{r}, t)) = (\mathbf{p}_0, v_0 \mathbf{p}_0 t) \ , \tag{12}$$

where $\mathbf{p}_0$ is a 2d vector of arbitrary direction and squared norm $|\mathbf{p}_0|^2 \equiv p_0^2 = -a/b$. It corresponds to an undeformed and uniformly polarized medium propagating with the velocity $v_0 \mathbf{p}_0$. We define the positively-oriented, orthonormal basis $(\mathbf{e}_\parallel, \mathbf{e}_\perp)$, where $\mathbf{e}_\parallel \equiv \mathbf{p}_0 / p_0$. Linearizing eqs. (10) & (11) around (12), and rescaling $\mathbf{r} \to \mathbf{r} - v_0 \mathbf{p}_0 t$, we get:

$$\partial_t \delta \mathbf{p} = -v_p p_0 \partial_\parallel \delta \mathbf{p} + \frac{2a}{\gamma |\mathbf{p}_0|^2} \mathbf{p}_0^{\otimes 2} \cdot \delta \mathbf{p} + \frac{\kappa}{\gamma} \Delta \delta \mathbf{p} + \frac{w}{2\gamma} \Delta \delta \mathbf{u} + \frac{w'}{\gamma} \nabla (\nabla \cdot \delta \mathbf{u}) \ , \tag{13}$$

$$\partial_t \delta \mathbf{u} = v_0 \delta \mathbf{p} + \frac{w}{2\Gamma} \Delta \delta \mathbf{p} + \frac{w'}{\Gamma} \nabla (\nabla \cdot \delta \mathbf{p}) + \frac{B}{2\Gamma} \Delta \delta \mathbf{u} + \frac{B'}{\Gamma} \nabla (\nabla \cdot \delta \mathbf{u}) \ . \tag{14}$$

We decompose the polarization $\mathbf{p} = \mathbf{p}_0 + \delta \mathbf{p}$ as $\mathbf{p} = (p_0 + \delta p)(\cos \theta \mathbf{e}_\parallel + \sin \theta \mathbf{e}_\perp)$, with $p_0 \equiv |\mathbf{p}_0|$. To linear order, $\delta \mathbf{p} \simeq \delta p \mathbf{e}_\parallel + p_0 \theta \mathbf{e}_\perp$. Denoting by $(\delta u^\parallel, \delta u^\perp)$ the components of $\delta \mathbf{u}$ in $(\mathbf{e}_\parallel, \mathbf{e}_\perp)$, the linearized dynamics (13) & (14) then read:

$$\partial_t \delta p = -v_p p_0 \partial_\parallel \delta p + \frac{2a}{\gamma} \delta p + \frac{\kappa}{\gamma} \Delta \delta p + \frac{w}{2\gamma} \Delta \delta u^\parallel + \frac{w'}{\gamma} \partial_\parallel (\partial_\parallel \delta u^\parallel + \partial_\perp \delta u^\perp) \ , \tag{15}$$

$$\partial_t \theta = -v_p p_0 \partial_\parallel \theta + \frac{\kappa}{\gamma} \Delta \theta + \frac{w}{2\gamma p_0} \Delta \delta u^\perp + \frac{w'}{\gamma p_0} \partial_\perp (\partial_\parallel \delta u^\parallel + \partial_\perp \delta u^\perp) \ , \tag{16}$$

$$\partial_t \delta u^\parallel = v_0 \delta p + \frac{w}{2\Gamma} \Delta \delta p + \frac{w'}{\Gamma} \partial_\parallel (\partial_\parallel \delta p + p_0 \partial_\perp \theta) + \frac{B}{2\Gamma} \Delta \delta u^\parallel + \frac{B'}{\Gamma} \partial_\parallel (\partial_\parallel \delta u^\parallel + \partial_\perp \delta u^\perp) \ , \tag{17}$$

$$\partial_t \delta u^\perp = v_0 p_0 \theta + \frac{w p_0}{2\Gamma} \Delta \theta + \frac{w'}{\Gamma} \partial_\perp (\partial_\parallel \delta p + p_0 \partial_\perp \theta) + \frac{B}{2\Gamma} \Delta \delta u^\perp + \frac{B'}{\Gamma} \partial_\perp (\partial_\parallel \delta u^\parallel + \partial_\perp \delta u^\perp) \ , \tag{18}$$

where $\partial_\parallel$ and $\partial_\perp$ respectively stand for the spatial derivatives along $\mathbf{e}_\parallel$ and $\mathbf{e}_\perp$.

The amplitude of the order parameter $\delta p$ relaxes to a value determined by the displacement fluctuations. To leading order in gradients, this yields,

$$\delta p \simeq \frac{w}{4a} \Delta \delta u^\parallel + \frac{w'}{2a} \partial_\parallel (\partial_\parallel \delta u^\parallel + \partial_\perp \delta u^\perp) \ . \tag{19}$$

Using this to eliminate $\delta p$ fluctuations we get

$$\partial_t \theta = -v_p p_0 \partial_\| \theta + \frac{\kappa}{\gamma} \Delta \theta + \frac{w}{2\gamma p_0} \Delta \delta u^\perp + \frac{w'}{\gamma p_0} \partial_\perp (\partial_\| \delta u^\| + \partial_\perp \delta u^\perp) \;, \quad (20)$$

$$\partial_t \delta u^\| = \frac{w' p_0}{\Gamma} \partial_\| \partial_\perp \theta + \left(\frac{B}{2\Gamma} + \frac{v_0 w}{4a}\right) \Delta \delta u^\| + \left(\frac{B'}{\Gamma} + \frac{v_0 w'}{2a}\right) \partial_\| (\partial_\| \delta u^\| + \partial_\perp \delta u^\perp) \;, \quad (21)$$

$$\partial_t \delta u^\perp = v_0 p_0 \theta + \frac{w p_0}{2\Gamma} \Delta \theta + \frac{w' p_0}{\Gamma} \partial_\perp^2 \theta + \frac{B}{2\Gamma} \Delta \delta u^\perp + \frac{B'}{\Gamma} \partial_\perp (\partial_\| \delta u^\| + \partial_\perp \delta u^\perp) \;. \quad (22)$$

To calculate the correlation functions of these quantities, we now introduce the Gaussian white noises $\lambda^\theta(\mathbf{r},t), \lambda^\|(\mathbf{r},t)$, and $\lambda^\perp(\mathbf{r},t)$, to the right-hand side of equations (20)-(22), whose statistics is given by $\langle \lambda^\alpha(\mathbf{r},t) \rangle = 0$ and $\langle \lambda^\alpha(\mathbf{r},t) \lambda^\beta(\mathbf{r}',t') \rangle = 2 D_{\alpha\beta} \delta(\mathbf{r}-\mathbf{r}') \delta(t-t')$, where $D_{\alpha\beta} \equiv D_\alpha \delta^{\alpha\beta}$ without any implicit summation over $\alpha$. Performing a spatial Fourier transform then yields

$$\partial_t \theta_k = -i v_p p_0 k_\| \theta_k - k^2 \frac{\kappa}{\gamma} \theta_k - k^2 \frac{w}{2\gamma p_0} \delta u_k^\perp - \frac{w'}{\gamma p_0} k_\perp (k_\| \delta u_k^\| + k_\perp \delta u_k^\perp) + \lambda_k^\theta \;, \quad (23)$$

$$\partial_t \delta u_k^\| = -\frac{w' p_0}{\Gamma} k_\| k_\perp \theta_k - \left(\frac{B}{2\Gamma} + \frac{v_0 w}{4a}\right) k^2 \delta u_k^\| - \left(\frac{B'}{\Gamma} + \frac{v_0 w'}{2a}\right) k_\| (k_\| \delta u_k^\| + k_\perp \delta u_k^\perp) + \lambda_k^\| \;, \quad (24)$$

$$\partial_t \delta u_k^\perp = v_0 p_0 \theta_k - \frac{w p_0}{2\Gamma} k^2 \theta_k - \frac{w' p_0}{\Gamma} k_\perp^2 \theta_k - \frac{B}{2\Gamma} k^2 \delta u_k^\perp - \frac{B'}{\Gamma} k_\perp (k_\| \delta u_k^\| + k_\perp \delta u_k^\perp) + \lambda_k^\perp \;, \quad (25)$$

where $k \equiv |\mathbf{k}|$ stands for the norm of the wave-vector $\mathbf{k} = k_\| \mathbf{e}_\| + k_\perp \mathbf{e}_\perp$, and $\langle \lambda_k^\alpha(t) \lambda_{k'}^\beta(t') \rangle = 2 D^{\alpha\beta} (2\pi)^2 \delta(k+k') \delta(t-t')$. To simplify the presentation, we ignore the advection of angular fluctuations by the polarisation, i.e., set $v_p = 0$ at this stage (and discuss its effect in a later subsection). This choice does not affect any of our qualitative discussions (except a particular feature of the mode structure that we discuss later). We introduce the vectors $X_k \equiv (\theta_k, \delta u_k^\|, \delta u_k^\perp)_k^\top$ and $\Lambda_k \equiv (\lambda_k^\theta, \lambda_k^\|, \lambda_k^\perp)^\top$, eqs (23)-(25) can be rewritten as:

$$\partial_t X_k = A_k X_k + \Lambda_k \quad (26)$$

where the matrix $A_k$ reads

$$A_k \equiv \begin{pmatrix} -\frac{\kappa}{\gamma} k^2 & -\frac{w'}{\gamma p_0} k_\| k_\perp & -\frac{w}{2\gamma p_0} k^2 - \frac{w'}{\gamma p_0} k_\perp^2 \\ -\frac{w' p_0}{\Gamma} k_\| k_\perp & \left(\frac{v_0 w}{4a} - \frac{B}{2\Gamma}\right) k^2 + \left(\frac{v_0 w'}{2a} - \frac{B'}{\Gamma}\right) k_\|^2 & \left(\frac{v_0 w'}{2a} - \frac{B'}{\Gamma}\right) k_\| k_\perp \\ v_0 p_0 - \frac{w p_0}{2\Gamma} k^2 - \frac{w' p_0}{\Gamma} k_\perp^2 & -\frac{B'}{\Gamma} k_\| k_\perp & -\frac{B}{2\Gamma} k^2 - \frac{B'}{\Gamma} k_\perp^2 \end{pmatrix} \;. \quad (27)$$

a. *Correlations.* Using Ito formula, we get

$$\partial_t (X_p^\alpha X_q^\beta) = X_p^\alpha \partial_t X_q^\beta + X_q^\beta \partial_t X_p^\alpha + 2 D^{\alpha\beta} (2\pi)^2 \delta(p+q) \;. \quad (28)$$

In steady state, this leads to

$$0 = \langle X_p^\alpha \partial_t X_q^\beta \rangle + \langle X_q^\beta \partial_t X_p^\alpha \rangle + 2 D^{\alpha\beta} (2\pi)^2 \delta(p+q) \quad (29)$$

$$= A_q^{\beta\mu} \langle X_p^\alpha X_q^\mu \rangle + A_p^{\alpha\mu} \langle X_q^\beta X_p^\mu \rangle + 2 D^{\alpha\beta} (2\pi)^2 \delta(p+q) \;, \quad (30)$$

which can be reformulated as

$$\mathcal{A}_{p,q}^{\alpha\beta\mu\nu} \langle X_p^\mu X_q^\nu \rangle = -8\pi^2 D^{\alpha\beta} \delta(p+q) \;, \quad (31)$$

with

$$\mathcal{A}_{p,q}^{\alpha\beta\mu\nu} = \delta^{\alpha\mu} A_q^{\beta\nu} + \delta^{\beta\nu} A_p^{\alpha\mu} \;. \quad (32)$$

By inverting the linear operator $\mathcal{A}_{k,-k}^{\alpha\beta\mu\nu}$, we obtain the correlation tensor

$$\langle X_k^\mu X_{-k}^\nu \rangle = -8\pi^2 \left[\mathcal{A}_{k,-k}^{-1}\right]^{\mu\nu\alpha\beta} D^{\alpha\beta} \;. \quad (33)$$

By computing $\mathcal{A}_{k,-k}^{-1}$ and only keeping the leading order term as $k \equiv |\mathbf{k}| \to 0$, we finally get the asymptotic large scale behavior of the correlators: They all scale as $1/k^2$, except for the transverse fluctuations of the displacement field, $\langle \delta u_k^\perp \delta u_{-k}^\perp \rangle = \langle X_k^\perp X_{-k}^\perp \rangle$, which scales as $1/k^4$. We discuss the consequence of this anomalously strong divergence later.

Using $\mathbf{v} = \partial_t \mathbf{u}$, together with the above correlators and eqs. (24) & (25), it can be straightforwardly deduced that

$$\langle \delta v_k^\parallel \delta v_{-k}^\parallel \rangle \sim \left( \frac{B'}{\Gamma} + \frac{v_0 w'}{2a} \right)^2 k_\parallel^2 k_\perp^2 \langle \delta u_k^\perp \delta u_{-k}^\perp \rangle , \qquad (34)$$

$$\langle \delta v_k^\perp \delta v_{-k}^\perp \rangle \sim v_0^2 p_0^2 \langle \theta_k \theta_{-k} \rangle , \qquad (35)$$

and hence that $\langle \delta v_k^\parallel \delta v_{-k}^\parallel \rangle$ and $\langle \delta v_k^\perp \delta v_{-k}^\perp \rangle$ respectively scale as a constant and as $1/k^2$. This shows that in direct space the fluctuations $\langle \delta v^\parallel \delta v^\parallel \rangle$ are short range (massive), while the fluctuations $\langle \delta v^\perp \delta v^\perp \rangle$ are algebraic (massless), in agrrement with experimental observations.

  b. *Number fluctuations.* From the correlators $\langle X_k^\alpha X_{-k}^\beta \rangle$ and eq. (5), we get that

$$\langle \delta \rho_k \delta \rho_{-k} \rangle \sim \rho_0^2 k_\perp^2 \langle \delta u_k^\perp \delta u_{-k}^\perp \rangle \qquad (36)$$

to leading order in $1/k$. This implies that $\langle \delta \rho_k \delta \rho_{-k} \rangle$ scales as $1/k^2$. This anomalously strong divergence—usually characteristic of a *broken-symmetry* variable and not a *conserved* one—is impossible in any equilibrium system (not at a critical point). Since number fluctuations is related to the correlations of $\delta \rho$ as

$$\langle \delta N^2 \rangle = L^2 \langle \delta \rho_k \delta \rho_{-k} \rangle_{k \to 0} , \qquad (37)$$

we conclude that

$$\langle \delta N^2 \rangle \sim L^4 \sim \langle N \rangle^2 , \qquad (38)$$

*i.e.* the systems undergoes giant number-fluctuations, as reported in the experimental graph of Figure 4-(j).

  c. *Eigenfrequencies.* Let us change to polar coordinates in Fourier space, *i.e.* $\mathbf{k} = (k \cos \phi, k \sin \phi)^\top$, where $\phi$ is the angle of $\mathbf{k}$ with respect to $\mathbf{e}_\parallel$. To second order in $k$, the first mode is purely diffusive with a dispersion relation :

$$\omega_d \simeq -ik^2 D(\phi) , \quad \text{with} \quad D(\phi) \equiv \frac{4a(B'w \cos^2 \phi + Bw' \sin^2 \phi)) + 2aBw - wv_0 \Gamma(w + 2w')}{4a\Gamma(w + 2w' \sin^2 \phi)} , \qquad (39)$$

while the other two frequencies also have real (propagating) parts:

$$\omega_\pm = \pm k \sqrt{\frac{v_0(w + 2w' \sin^2 \phi)}{2\gamma}} - ik^2 \widetilde{D}(\phi) , \qquad (40)$$

with

$$\widetilde{D}(\phi) = \frac{B\gamma + 2\Gamma\kappa}{4\gamma\Gamma} + \frac{aB'(2w' + w) - w'^2 v_0 \Gamma \cos^2 \phi}{2a\Gamma(w + 2w' \sin^2 \phi)} \sin^2 \phi . \qquad (41)$$

Despite the inertialess nature of the dynamics, the system displays propagating modes, a widespread feature in active systems, which is why activity often mimics the role of inertia. These propagating modes (40) & (41) possess a front-back symmetry with respect to the direction $\mathbf{p}_0$ of the solid motion, a property they inherit from the linearized dynamics.

Even though the components $\theta, \delta u^\parallel$, and $\delta u^\perp$ are generically all entangled in each eigenmode, the diffusive mode is mainly supported by $\delta u^\parallel$, while the propagating modes are essentially along $\theta$ and $\delta u^\perp$. This fact, which is only approximately true in general, becomes exact at least in two situations: when we consider only wavevectors that are aligned with the direction of the solid motion and in the simplified model presented in section III, where the solid is assumed to be incompressible.

Interestingly, the speed of the propagating modes is maximum in the transverse direction. In addition, all the parameters of the model being positive, except $a$ which is negative, it can be easily shown that $\widetilde{D}(\phi) \geq \frac{B\gamma + 2\Gamma\kappa}{4\gamma\Gamma}$, with equality iff $\phi = n\pi, n \in \mathbb{Z}$. In other words, the propagating modes are minimally damped in the direction of the solid motion. Despite its noteworthy accordance with the experiments, where only modes that propagate along $\mathbf{p}_0$ are observed, this effect would have been absent if we had considered a more general anisotropic free energy density for $\mathbf{p}$ allowing for distortions along and transverse to the ordering direction to have different Frank elasticities.

*d. The effect of self-advection on eigenmodes.* In this paragraph, we analyse the consequences of the self-advection of polarisation in Eq. (23) for the eigenfrequencies. We only display the effect of this to lowest order in wavenumbers.

As in the case without self-advection, one of the three modes remains diffusive while the front-back symmetry of the other two modes (40) with respect to the direction of solid motion is broken leading to the eigenfrequencies

$$\omega_\pm = k \left[ \frac{v_p}{2} p_0 \cos\phi \pm \sqrt{\frac{(v_p p_0 \cos\phi)^2}{4} + \frac{v_0(w + 2w' \sin^2\phi)}{2\gamma}} \right] + \mathcal{O}(k^2) . \tag{42}$$

This asymmetry in the direction of propagation of longitudinal waves is consistent with observations, which report only backward propagating waves.

## III. A SIMPLIFIED MODEL

To elucidate which couplings are crucial to get the divergence of $\langle \delta u_k^\perp \delta u_{-k}^\perp \rangle$ in $1/k^4$ and allow for simpler, fully explicit analytical expressions, we start by assuming that $B' = w' = 0$ (or that the solid is incompressible). We also take $v_p = 0$ as this doesn't affect the scaling of the correlator. In this case, the dynamics of $\delta u^\parallel$ decouples from that of $\theta$ and $\delta u^\perp$, while the latter two fields satisfy :

$$\partial_t \theta = \frac{\kappa}{\gamma} \Delta \theta + \frac{w}{2\gamma p_0} \Delta \delta u^\perp + \lambda^\theta , \tag{43}$$

$$\partial_t \delta u^\perp = v_0 p_0 \theta + \frac{w p_0}{2\Gamma} \Delta \theta + \frac{B}{2\Gamma} \Delta \delta u^\perp + \lambda^\perp . \tag{44}$$

Taking a space-time Fourier transform gives

$$\left( -i\omega + \frac{\kappa}{\gamma} k^2 \right) \theta_{k,\omega} = -\frac{w}{2\gamma p_0} k^2 \delta u_{k,\omega}^\perp + \lambda_{k,\omega}^\theta , \tag{45}$$

$$\left( -i\omega + \frac{B}{2\Gamma} k^2 \right) \delta u_{k,\omega}^\perp = p_0 \left( v_0 - \frac{w}{2\Gamma} k^2 \right) \theta_{k,\omega} + \lambda_{k,\omega}^\perp . \tag{46}$$

Using eq. (45), we can express $\theta_{k,\omega}$ as a function of $\delta u_{k,\omega}^\perp$ and $\lambda_{k,\omega}^\theta$. Reinjecting the result in eq. (46) and rearranging the terms, we get

$$\delta u_{k,\omega}^\perp = \frac{p_0 \left( v_0 - \frac{w}{2\Gamma} k^2 \right) \lambda_{k,\omega}^\theta + \left( i\omega + \frac{\kappa}{\gamma} k^2 \right) \lambda_{k,\omega}^\perp}{-\omega^2 - i\omega \left( \frac{\kappa}{\gamma} + \frac{B}{2\Gamma} \right) k^2 + \frac{w v_0}{2\gamma} k^2 + \left( \frac{B\kappa}{2\Gamma\gamma} - \frac{w^2}{4\gamma\Gamma} \right) k^4} . \tag{47}$$

The denominator of this equation, which we denote by $P(\omega)$, has the following pair of roots for sufficiently small $k$:

$$\omega_\pm = -\frac{i}{2} \left( \frac{\kappa}{\gamma} + \frac{B}{2\Gamma} \right) k^2 \pm \frac{1}{2} \left[ \frac{2wv_0}{\gamma} k^2 + k^4 \left( \frac{B\kappa}{\Gamma\gamma} - \frac{w^2}{\Gamma\gamma} - \frac{\kappa^2}{\gamma^2} - \frac{B^2}{4\Gamma^2} \right) \right]^{1/2} . \tag{48}$$

Multiplying each side of (47) by its complex conjugate and using $\langle \lambda_{k,\omega}^\alpha \lambda_{k',\omega'}^\beta \rangle = 2D_{\alpha\beta} (2\pi)^3 \delta(\mathbf{k} + \mathbf{k}') \delta(\omega + \omega')$, with $D_{\alpha\beta} \propto \delta_{\alpha\beta}$, we get

$$\langle |\delta u_{k,\omega}^\perp|^2 \rangle = 16\pi^3 \frac{p_0^2 \left( v_0 - \frac{w}{2\Gamma} k^2 \right)^2 D_\theta + \left( \omega^2 + \frac{\kappa^2}{\gamma^2} k^4 \right) D_u}{(\omega - \omega_-)(\omega - \omega_+)(\omega - \omega_-^*)(\omega - \omega_+^*)} , \tag{49}$$

where the asterisk stands for complex conjugation. To get the equal-time correlator, we then integrate with respect to $\omega$:

$$\langle |\delta u_k^\perp|^2 \rangle = \frac{1}{(2\pi)^2} \int \langle |\delta u_{k,\omega}^\perp|^2 \rangle d\omega . \tag{50}$$

Computing this integral with contour integration in the complex plane, we see that the dominant as $k \to 0$ is simply

$$\langle |\delta u_k^\perp|^2 \rangle \simeq 4\pi \int \frac{p_0^2 v_0^2 D_\theta}{(\omega - \omega_-)(\omega - \omega_+)(\omega - \omega_-^*)(\omega - \omega_+^*)} d\omega . \tag{51}$$

Using residue theorem we finally get

$$\left\langle |\delta u^\perp_{k,\omega}|^2 \right\rangle \simeq -2\pi p_0^2 v_0^2 D_\theta \frac{2i\pi}{(\omega_+ + \omega_-)\omega_+\omega_-} \tag{52}$$

$$= \frac{4\pi^2 p_0^2 v_0^2 D_\theta}{k^2 \left[\frac{\kappa}{\gamma} + \frac{B}{2\Gamma}\right] \left[\frac{k^4}{4}\left(\frac{\kappa}{\gamma} + \frac{B}{2\Gamma}\right)^2 + \frac{wv_0}{2\gamma}k^2 + \frac{k^4}{4}\left(\frac{B\kappa}{\Gamma\gamma} - \frac{w^2}{\Gamma\gamma} - \frac{\kappa^2}{\gamma^2} - \frac{B^2}{4\Gamma^2}\right)\right]} \tag{53}$$

$$\simeq \frac{16\pi^2 p_0^2 v_0 D_\theta \gamma^2 \Gamma}{k^4 w(2\Gamma\kappa + B\gamma)} \tag{54}$$

Let us now identify what could have been neglected from the initial dynamics while still keeping the $1/k^4$ scaling of $\left\langle |\delta u^\perp_{k,\omega}|^2 \right\rangle$. While approximating eqs. (49) & (50) by (51), we have seen that $\lambda^\perp$ and the $\Delta\theta$ in eq. (44) could have both been neglected from the start. Further, by tracing back to eq. (48) and $P(\omega)$ the terms we neglected between eqs. (53) & (54), we see that we could have also set either $B$ or $\kappa$ to zero (but not both).

We now comment on the strong small $k$ divergence of $\left\langle |\delta u^\perp_{k,\omega}|^2 \right\rangle \sim 1/k^4$. In two-dimensional real space, this implies that displacement fluctuations $\left\langle |\delta u^\perp(\mathbf{r},t)|^2 \right\rangle \sim L^2$, where $L$ is the system size. This is in contrast with active polar elastomer in the hydrodynamic limit (i.e., one in which the effect of $c$ is accounted for) in which both longitudinal and transverse displacement fluctuations diverge logarithmically. This strong divergence further implies that *strain* fluctuations $\sim \nabla \mathbf{u}$ diverge logarithmically with system size. This implies that the solid only has short-range order and, at large scales, the solid structure is washed away by fluctuations, which in practice would lead to rupture or melting depending on the choice of rheological model at large strain. Of note, this assumes that $c$ does not affect the dynamics, which is expected to hold at intermediate scales, but not at arbitrarily large scales.